\documentclass[%
 reprint,
 amsmath,amssymb,
 aps,
pra,
twocolumn,
]{revtex4-2}

\usepackage{graphicx}
\usepackage{dcolumn}
\usepackage{bm}
\usepackage{amsmath}
\usepackage{amssymb}
\usepackage{xcolor}
\usepackage{graphicx}
\usepackage{todonotes}
\usepackage{siunitx}
\usepackage{comment}

\begin{document}


\title{Urban topology and dynamics can assess green areas importance}

\author{Jacopo Moi}
 \affiliation{
 DSMN and ECLT, Ca'Foscari University of Venice\\
 Fondazione Futuro delle citt\`a, Florence, Italy
 }
 \author{Leonardo Chiesi}%
\affiliation{ 
University of Florence, Florence Italy\\
Fondazione Futuro delle citt\`a, Florence, Italy
}%
\author{Gherardo Chirici}%
 \email{Gherardo.Chirici@unifi.it}
\affiliation{ 
University of Florence, Florence Italy\\
Fondazione Futuro delle citt\`a, Florence, Italy
}%
\author{Saverio Francini}
\affiliation{University of Florence, Florence Italy\\
Fondazione Futuro delle citt\`a, Florence, Italy}
\author{Costanza Borghi}
\affiliation{University of Florence, Florence Italy\\
Fondazione Futuro delle citt\`a, Florence, Italy}

\author{Paolo Costa}
\affiliation{University of Florence, Florence Italy\\
Fondazione Futuro delle citt\`a, Florence, Italy}
\author{Bianca Galmarini}
\affiliation{University of Florence, Florence Italy\\
Fondazione Futuro delle citt\`a, Florence, Italy}

\author{Guido Caldarelli}%
 \email{Guido.Caldarelli@unive.it}
\affiliation{ 
DSMN and ECLT, Ca'Foscari University of Venice, Italy\\
Institute of Complex Systems CNR (ISC CNR), Depart. Phys. "Sapienza" University Rome, Italy\\
London Institute for Mathematical Sciences, Royal Institution, London UK
}%

\date{\today}

\begin{abstract}
Green areas are a crucial element in evolution of a city, contributing to improve citizens' life, to reduce effects of climate change, and to make possible the survival of other species in urban areas.  Unfortunately, the above effects are difficult to assess quantitatively for regulators, stakeholders and experts, making troublesome the planning of city development. Here we present a method to estimate the impact of these areas in the city life based on the network topology of the city itself and on a simple model of dynamics on this structure. Movements between various areas of the city are simulated by means of an agent-based biased-diffusion process where citizens try to reach the nearest Public Green Area (PGA) from their position and the model is fed with real data about the density of populations in the cases of study. Firstly, we define a centrality measure of PGA's based on average farness measured on the city network; this approach outperforms information based on the simple topology. We then improve this quantity by taking into account the occupation of PGA's, thereby providing a quantitative measure of PGA usage for regulators. 
\end{abstract}

\maketitle

\section*{Introduction}
Urban green spaces, which encompass parks, gardens, urban forests, and other forms of vegetation, provide cities with a diverse array of environmental, climate, economic, social, and health benefits. From an ecological point of view, the extension of urban green infrastructures,  improves biodiversity and ensures the proper functioning of ecological systems, fostering the prosperity of various animal species \cite{marc2017evolution}. Green areas also provide a series of ecosystem services (such as air purification, carbon sequestration, urban cooling, runoff control, noise reduction, and habitat maintenance \cite{jax2013ecosystem}) essential for city life. Finally. green spaces improve city resilience, particularly against the adverse effects of climate change, and enhance the overall quality of the urban environment. The economic significance of green spaces is evident as well, as greening is frequently linked to higher home prices and increased property values, and serves as an effective strategy for branding and attracting international capital investments \cite{anguelovsky2019why}.
However, beyond the ecological and the economic realm, the value of green spaces and particularly the Public Green Areas (PGA) becomes even more important when viewed from the citizens’ perspective. A robust and ecologically healthy natural component is crucial for health and well-being. Enhancing environmental conditions positively impacts life, particularly in cities where pollution, noise, and heat pose significant challenges. Furthermore, it is now clear how direct contact with nature is a vital condition to achieve an healthy urban life reducing stress, anxiety, and depression, as well as reduced risks of obesity, cardiovascular diseases, and respiratory illnesses \cite{barboza2021green,bratman2019nature,harting2014health,kuo2015contact,triguero2017natural}.
Among less immediate benefits, exposure to nature facilitates personal growth and learning, favours prosocial behaviours and cooperation \cite{goldy2019social}, enhances social cohesion \cite{jennings2019cohesion} and environmental awareness.
The structure, size, connectivity, and biodiversity of green spaces are determinants for the provision of benefits \cite{bratman2019nature}. However, health and social benefits also hinge on the dose of nature experience \cite{shanahan2016benefits}, thus its frequency, duration, and recurrence. The variation in these factors raises equity concerns, as differences in access among different segments of the population can compromise the fair distribution of health benefits. Consequently, distributive and accessibility considerations regarding green spaces in cities emerge as indispensable components for informing a just green planning. We present here a metric based on statistical physics framework to assess the relevance of PGAs in city life. The metric takes into account position and shape of areas with respect to the rest of the city and the facility to access these areas by the population. A way to assess quantitatively the accessibility of these areas is to measure their centrality in the city structure.
The approach of Statistical Physics is particularly suitable in this case, given the large sizes and the strong heterogeneity of the urban systems~\cite{barthelemy2019statistical}.
More particularly complexity science~\cite{batty2005cities,verbavatz2020growth} and concepts developed in network theory\cite{caldarelli2007scale}, as the centrality mentioned above, are the most effective ways to achieve a green area metric. Indeed, by reducing the complexity of cities\cite{caldarelli2023role} to a web of city blocks in the form of a network we can identify even social patterns of urban city life that, when quantified, appear to follow clear mathematical laws\cite{depersin2018scaling,gallotti2021unraveling}.
Recent research on human interactions in cities embraced all these ideas, trying to produce models that, starting from different data, create comprehensive multilevel networks\cite{gallotti2021complex} that can help understand the phenomenon better than a digital twin\cite{caldarelli2023role}.
Indeed, with such a macroscopic level of description, it is possible to detect a series of “universal” patterns emerging in cities worldwide\cite{facchini2021urban}, indicating that the size of a city – measured by its population – is a primary determinant of its characteristics. Through this quantity, it is possible to assess the importance of the various zones of the city and the other relative properties as the connectivity, the clustering, the assortativity effects\cite{catanzaro2004social}, {\em etc}. 
In detail, we consider the network structures obtained by Voronoi tessellation for three Italian cities of different sizes and different populations (high, medium, medium/small): Milan, Florence and Mestre, respectively. On these structures, we evaluate the accessibility of  PGAs at the city block level by utilizing both a traditional network centrality measure and modified versions that account for spatial interaction and population density. On these areas, we first perform a centrality analysis, and then, with the help of a simple dynamical model, we obtain a computational measure of the usability of such areas. 
\section*{Methods}
\subsection{Voronoi Tassellation and Networks}
We base our analysis on weighted spatial networks created through the tessellation \cite{voronoi1908nouvelles} of three Italian cities: Mestre (Venice), Florence, and Milan. The weighted spatial network, denoted as $G(V, E)$, comprises nodes ($V$) representing centroids, and weighted edges ($E$) capturing their interactions. Nodes are defined as $V = {v_1, v_2, ..., v_N}$, and edges are represented as $E = {(v_i, v_j, w_{ij})}$, with $w_{ij}$ representing the Euclidean distance between adjacent nodes $v_i$ and $v_j$. In our network representation, we have two types of nodes: city blocks (denoted as $b$) with associated population data from the census\cite{istat2011census}, and Public Green Areas (PGA, denoted as $p$). We take this data from ``OpenStreetMap''\cite{OpenStreetMap}.
Each city displays distinct features in population density and urban layout. City of Mestre has a concentrated, recent development, while Florence shows an uneven distribution of population and city blocks some of which with hundreds of years of story. Milan, a densely populated metropolis, exhibits a more regular urban structure with ancient and new buildings and with varying population densities.
To evaluate the importance of block nodes relative to the set of PGAs nodes, we employed various measure of network centrality. 
In particular, we shall focus on the idea that the probability of performing a walk between a block and a PGA follows a well-defined probability distribution that depends on both network topology and edge length as cost. In the following, we shall look for the ensemble of ``low cost" paths, from any city block to the PGA's. Furthermore, PGA nodes that are the target of these walks will be modeled in the first instance as 'sink' of infinite capacity and in the second instance with a finite capacity given by their walkable area.

\subsection{Taking into account the social network}
Cities, in general, are characterized by different population densities per block, which inevitably play a determining factor in characterizing the use of and accessibility to certain services. This is well known even before the advent of complex networks, from the principle of human behaviours\cite{zipf1949human}, to models explicitly pointing out the role of distance in social interactions\cite{zipf1946gravity}, to the advantageous concentration of diverse businesses and individuals in urban areas promoting economic vitality through closeness~\cite{jacobs1961death} over time. All the possible social contacts and relationships with transportation and supply networks, concur in creating a series of different co-evolving structures describing the evolution of life in cities.  The usability of PGA's is inherently tied to their physical proximity to the population, a factor that, in turn, significantly influences their potential use. In other terms, the quantity of population physically served by each green space is a determinant for rating its role in service providing and its likelihood to use. According to the least effort hypothesis \cite{zipf1949human}, proximity has a pivotal role in shaping behaviours and choices, and consequently, social interactions. In the present study, we advance a dynamic network-based model that posits people density and proximity as primary determinants of agency. At the same time, we acknowledge the intricate interplay of numerous other variables, here unexplored and more difficult to model, influencing the use of public green areas, including design, governance mode, maintenance level, biodiversity, quality, dimension, but also the non-normativity and functional indeterminacy\cite{chiesi2022small}.
\subsection{Probability of paths to PGAs nodes}\label{method:bagofpath}
The accessibility to Public Green Areas (PGAs) is defined as the probability of a block resident's access to PGAs, primarily based on the distance between the block and PGAs. To derive a probability distribution for each block, we employed the bag-of-path approach\cite{francoisse2017bag}, which employs the principles of random walk theory within network analysis.
A random walk on a network is a stochastic process that starts from a given vertex $v$, and then selects one of its neighbors according to a probability distribution.
A naturally defined distribution arises by imposing a uniformly likely passage on the edges. This means to consider the degree of the node $ P_{v\rightarrow u}=\frac{1}{\text{Deg}(v)}$, if and only if $u,v$ are connected. The latter relationship can be calculated for each pair of nodes $u,v$ and stored in a transition matrix $\mathbf{P}_{\text{ref}}$. A typical process in networks is when specific nodes are made {\em{adsorbing}}, causing random walks from node \(i\) to halt indefinitely at node \(j_{\text{ads}}\). This properties can be modelled by placing $p_{j_{ads},j_{ads}}=1, p_{j_{ads},i}=0$ in the transition matrix.
In this case, a quantity of interest is the expected number of times that a walk from any node $i$ reaches one of the absorbing nodes ${j_{\text{ads}}}$. This is computed for each starting node by summing all possible paths of length \(t\), accounting for the adsorbing property of selected nodes\cite{masuda2017random}:
\begin{equation}
    \mathbf{Z}=\sum_{t=0}^{\infty}\mathbf{P}_{\text{ref}}^{\:t}=(\mathbf{I}-\mathbf{P}_{\text{ref}})^{-1}\label{eq:rnd:fund_matrix}
\end{equation}
The $\mathbf{Z}$ matrix is called the {\em{fundamental matrix}} and each entry $Z_{ij}$ contains the expected number of times that a random walk starting in $i$ visits one node $j$ before being adsorbed.
Since edge weights represent Euclidean distance in our spatial networks, we aim to calculate the probability of observing a path from a block node $b$ to one of the PGA nodes, treating them as absorbing, with path cost influenced by edge weights. The objective is to obtain a probability distribution where longer paths are penalized in terms of cost. 
Assuming path costs equal path length, a statistical ensemble minimizing the expected total cost \(\mathbb{E}[c(\mathcal{P})]\) can be derived \cite{francoisse2017bag,Saerens2009Randomized}. This leads to the identification of a new transition matrix, incorporating the natural matrix \(\mathbf{P_{\text{ref}}}\) and exponentially attenuated edge costs (a sort of Boltzmann's weight). This represents a system of nodes considered to be partially absorptive {{\em{i.e.}} the probability is not conserved on paths}:
\begin{equation}
    \mathbf{W}=\mathbf{P_{\text{ref}}}\circ e^{-\gamma\mathbf{C}}
\end{equation}

Here 
\begin{itemize}
\item the symbol $\circ$ corresponds to the Hadamard's matrix product\cite{hadamard1900product}  where entries of the matrices are multiplied cell by cell;
\item $\gamma$ (corresponding to an inverse temperature) is a parameter controlling the ensemble bias toward shortest paths; 
\item $\mathbf{C}$ is the cost matrix equivalent to the network's weight matrix; the fundamental matrix $\mathbf{Z}$ is derived from $\mathbf{W}$ as in Eq.\eqref{eq:rnd:fund_matrix} and it naturally defines the partition function of this random walk system. 
\end{itemize}

Leveraging this property allows defining the probability of observing a direct path from a block \(b\) to a PGA node \(p\) (but not {\em vice versa}) by first diagonalizing 
\begin{equation}
\mathbf{Z^h}=\frac {\mathbf{Z}} {\text{Diag}(\mathbf{Z})}
\end{equation}
and then row-normalizing:
\begin{equation}
    \mathbf{Z^{\textbf{PGA}}}_{i,j}=(\mathcal{P}(b=i,p=j))=\frac{Z^{h}_{i,j}}{\sum_{k \in {\text{PGA}}}Z^h_{i,k}}
\end{equation}
$P(\mathcal{P}(b,p))$ thus corresponds to the probability of reaching the PGA $p$ (made implicitly adsorbing) from block $b$ without being killed during the walk.
The matrix $\mathbf{Z^{\textbf{PGA}}}$ collects the probabilities calculated above for every blocks.

\subsection{Network centrality}
Network centrality\cite{newman2010network} measures the ``importance'' of a given vertex with respect of the others in the network. Various centrality measures are possible, according to what we consider important.
The Farness Centrality of a node in a graph is a measure of the total distance of a given node to a subset of nodes in the graph. 
It is defined as:
\begin{equation}
    C_f(b) = \sum_{ p \in \{\text{PGA}\}} d_{SP}(b, p)
\end{equation}
Where $d_{SP}$ is the weighted distance ({\em i.e.} the shortest path) between nodes $b,p$.
Visually, this calculated metric is akin to computing the sum of distances within a circle that stretches to encompass the farthest nodes, treating each path equally as contributing to the node's centrality. However, in an urban context, it is not necessarily taken for granted that the actual contribution of the shortest-path length is proportional to its distance. 
Consider, for example, the case where a block has equidistant accesses to a set of PGAs ({\em i.e.} PGAs arranged on a circumference); within this metric, this is the most favoured block. In contrast, a block close to the circumference will experience a greater Farness but a larger set of short paths toward near PGAs, which turn out to be more likely to be crossed.
This observation suggests that, given a distance-dependent probability distribution $P(d_{SP}(b,v))$ of observing a path from block $b$ to the set of PGAs, an Average Farness defined as follows is more informative:

\begin{equation}
    \langle C_f(b) \rangle = \sum_{ v \in \{\text{PGA}\}} d_{SP}(b, v)P(d_{SP}(b,v))
    \label{eq:avgfarness:general}
\end{equation}

Where $P(d_{SP}(b,v))$ is a suitable probability distribution that describes the 'importance' of the paths $b,v$. 
In so doing, we will obtain values of centrality that range from $0$ to the largest possible distance. For further applications, we might need a normalized version $C^N_f$ of this class of centralities. In general, we shall transform those quantities by using the following formula (Max-Min)
\begin{equation}
     C^N_f  = \frac{\text{max}(C_f)-C_f}{{\text{max}(C_f) - \text{min}(C_f)}}
     \label{eq:minmax}
\end{equation}

\subsubsection{Average farness involving low-cost paths}
Here is defined the Average Farness of a block $b$ toward the whole set of PGAs,
by resorting to the directed hitting-paths probability matrix $\mathbf{Z^{\textbf{PGA}}}$:
\begin{equation}
\begin{split}
    \langle C_f(b) \rangle = 
    \sum_{ v \in \{\text{PGA}\}} d_{SP}(b, v)P(d_{SP}(b,v))&=\\
    \sum_{ v \in \{\text{PGA}\}} d_{SP}(b, v)Z^{\text{PGA}}_{b,v} \label{eq:avgh}
\end{split}
\end{equation}
This measure describes how much a block is locally far from the rest of PGAs in accordance with Eq.\eqref{eq:avgfarness:general}. This is fair because a walker will experience only the situation from his starting block. A conceptual limitation of this metric is based on defining PGA nodes as infinitely adsorbing and thus thought of as inherently having infinite capacity. As introduced, PGAs possess a finite walkable area, which gives rise to access competition among residents.
In this case, the quality of a block is best described by an occupancy probability distribution $P_{\text{PGA}}^{\:b}$ regarding the probability that residents from the block $b$ will access the set of PGA nodes, given the different path probability toward these.

\subsubsection{Average farness from PGAs occupancy distribution}
This centrality is based on the PGA occupancy probability distribution $P_{PGA}^b$ per block $b$. This distribution simply represents the frequency of residents of block $b$ who have accessed the i-$th$ PGA. 
This distribution is inserted into the Average Farness expression:
\begin{equation}
\begin{split}
    \langle C_f(b) \rangle_{PGA} = 
    \sum_{ v \in \{\text{PGA}\}} d_{SP}(b, v)P_{\text{PGA}}^{\;b}(v) \label{eq:avghag}
\end{split}
\end{equation}

To model the distribution, we presumed a latent competition among city residents for access to PGAs. The distribution is then sampled from the collective outcomes of independent agents seeking low-cost paths to PGAs.
Below the agent model is introduced and details onto the calculation of $P_\text{PGA}$ are then provided.
\subsection{Agent-based model}\label{subsec:agmod}
We simulated the behaviour of the population (agents) with a process akin to diffusion. The diffusion is not an ordinary random diffusion (where people would wander around with no direction), rather in this agent-based model, each agent acts as a biased random walker, aiming to stop at a PGA capable of accommodating the agent's searched area, denoted as $f_{mq}$. From this parameter we then obtain $MaxOcc_{i} = \frac{Area(PGA)}{f_{mq}}$, which represents the maximum number of agents that can be hosted by the $i$-th PGA in the city. Agents possess a limited amount of energy ({\em{i.e.} total length walked}) expressed in meters. During the walk, each agent incurs an energy loss $\Delta E$ equal to the length of the crossed edge.
There are three possible final states for an agent: (i) staying in a PGA node, (ii) reaching a PGA without staying there, or (iii) running out of energy without reaching a PGA. As introduced, agents compete for occupation of the PGAs. When an agent reaches a PGA, the decision-making process involves testing $\mathcal{X} > Occ_{i}$, where $Occ$ is the actual occupancy within a PGA expressed as the ratio of the current number of agents who stopped in the PGA to the maximum occupancy. We decided to evaluate the case where $\mathcal{X}$ follows a uniform distribution $\mathcal{X}\sim U(0,1)$. If the outcome is positive, the agent stops its walk in the PGA (i). Otherwise, it continues the walk from the PGA node (ii). In the case of state (iii), the agent restarts the walk from its original node.
In this way, the agents follow low-cost paths leading to PGAs, determined by the optimization of a 'reward' function $\phi_i$ for each block $i$.
The $\phi$ is derived directly by the matrix $\mathbf{Z}$ in the following way: 
\begin{equation}
\phi_i=\mathbf{e_{i}}^{\text{T}}\mathbf{Z}\mathbf{e}_{\{v\}}
\end{equation}
Where $\mathbf{e_i}$ is the vector indexing the $i$-th block.
The reward function $\phi$ is the sum of the expected number of times that the $i$-th block can potentially visit the set of PGA nodes.
In that sense, nodes that possess a high reward have a better quality in terms of low-cost paths than ones with a lower one.
The $\phi$ function increase along paths leading to PGAs, with PGA nodes possessing the highest $\phi$ across the network. Under this approach, each node in the network is informed about its neighborhood quality of PGAs in terms of distances.
The transition probability employs the Monte Carlo sampling strategy:
\begin{equation}
    P_{i \rightarrow j}=\frac{e^{-\beta\Delta\phi_{ij}}}{\sum_{k \in N(i)}e^{-\beta\Delta\phi_{ki}}} \label{eq:prob}
\end{equation}
where $N(i)$ is the neighbors set of the node $i$. 
An agent walk thus interpolates between a random walk and a walk that follows low-cost paths but without a PGA chosen in advance. In any case, as we shall see, green areas close to a block in terms of distance are more likely to be reached by walkers, and conversely, occupation of distant ones is a rare event triggered by overcrowding of nearby ones.
To account for population density disparities within the cities, agents are further distinguished by unique starting nodes allocated in proportion to the population distribution across the blocks.
Each simulation was repeated by varying the $f_{mq}$ parameter across a range of values. This approach enables an investigation of agent dynamics across a spectrum, transitioning from scenarios abundant in walkable green areas to those with a scarcity.
At the end of each simulation, agents' paths were systematically collected to extract the 
occupancy probability distribution $P_{PGA}(b, f_{mq})$ per block $b$, which is dependent from the parameter $f_{mq}$. To recover a comprehensive distribution, the marginal distribution was then recovered:
\begin{equation}
    P_{\text{PGA}}^{b}=\sum_{f_{mq}}P_{\text{PGA}}(b, f_{mq})
\end{equation}
\paragraph{Cumulative density}
PGAs act as sinks that absorb a quantity of material (i.e., agents) from blocks at different distances.
To understand the relation between population density and PGAs, a simple metric was defined:
\begin{equation}
    \rho_{>}(m)=\frac{1}{N_{a}}
\sum_{p \in \{\text{PGA}\}} \sum_{j \in \{V_{b}^{p}\} } \mathbf{1}\left\{d_{\text{SP}}(p,j)\leq m\right\}\label{eq:cumdens}
\end{equation}
Where $N_{a}$ stands for the total number of agents, $V_{b}^{i}$ the set of blocks from which the PGA $j$ has sourced a number of occupants, and $d_{SP}$ the weighted shortest distance between the PGA and the $j$-th block. This metric is normalized between 0 and 1, so it allows us to explore the Cumulative density of inhabitants that reached PGAs at different distances with respect to the $i$-th bins $m$.
\begin{figure}[!ht]
    \centering
    \includegraphics[width=0.48\textwidth]{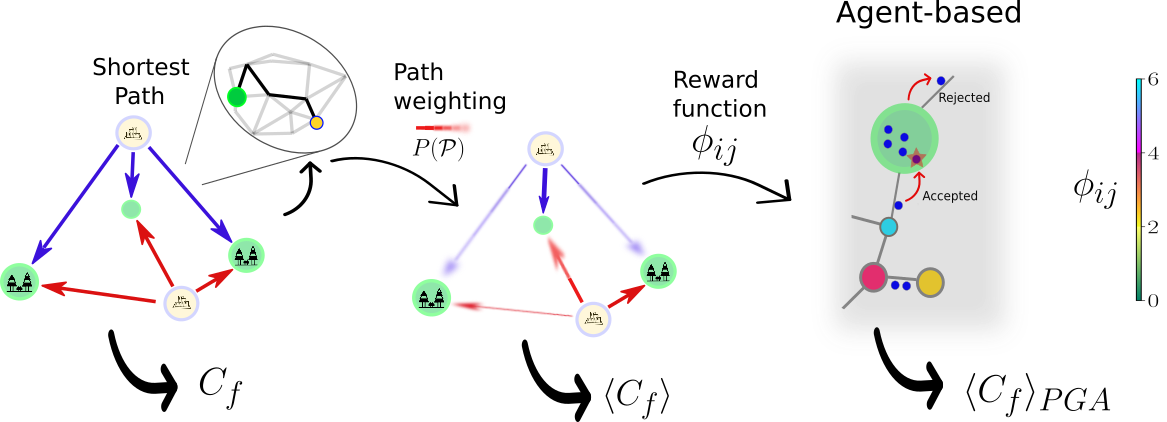}
    \caption{Main elements of the method devised in this work. From left: farness centrality calculated from shortest paths; average farness obtained by path weighting (bag-of-path approach); reward function and agent-based simulations.}
    \label{fig:misc:method}
\end{figure}
\section{Results and Discussion}
\subsection{Average Farness centrality}\label{subsec:avfarn}
Initially, the hitting path probability matrix $\mathbf{Z^{\mathbf{PGA}}}$ was derived for the three specified cities. The $\gamma$ parameter was thus set to $\gamma=\num{8.3e-4}$, which corresponds to the ensemble of paths given the reference 15-minutes distance of $\SI{1200}{m}$. 

\begin{figure}[!ht]
    \centering
    \begin{minipage}[b]{0.35\textwidth}
        
        \label{fig:cent:firenze}
        \includegraphics[trim={0.5cm 0.1cm 0.5cm 0cm},clip,width=\textwidth]{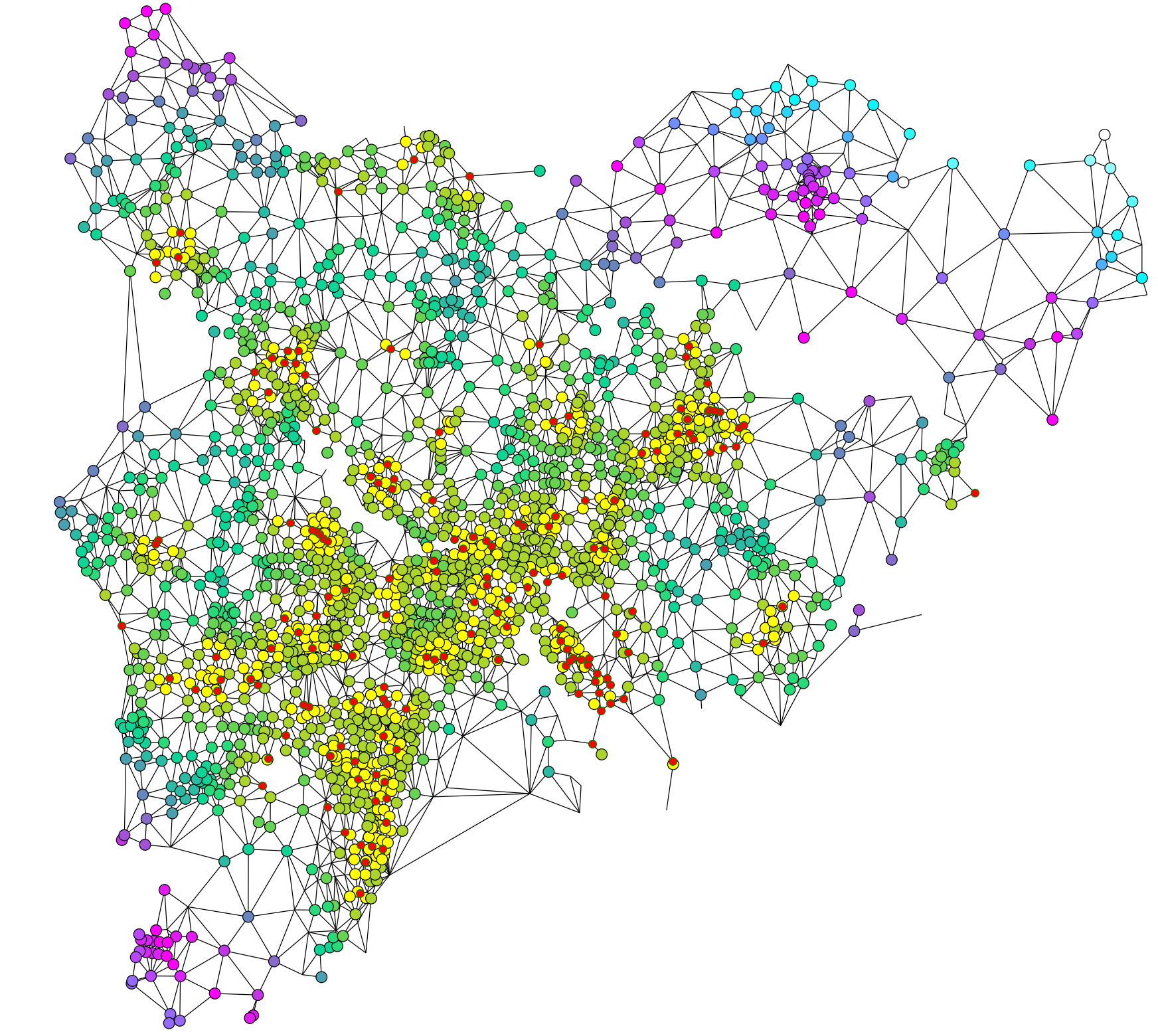}\label{fig:cent:avfarn:mestre}
        (a) Mestre\\
    \end{minipage}
    \begin{minipage}[b]{0.35\textwidth}
        
        \label{fig:cent:milan}
        \includegraphics[trim={0cm 0.1cm 0cm 0cm},clip,width=\textwidth]{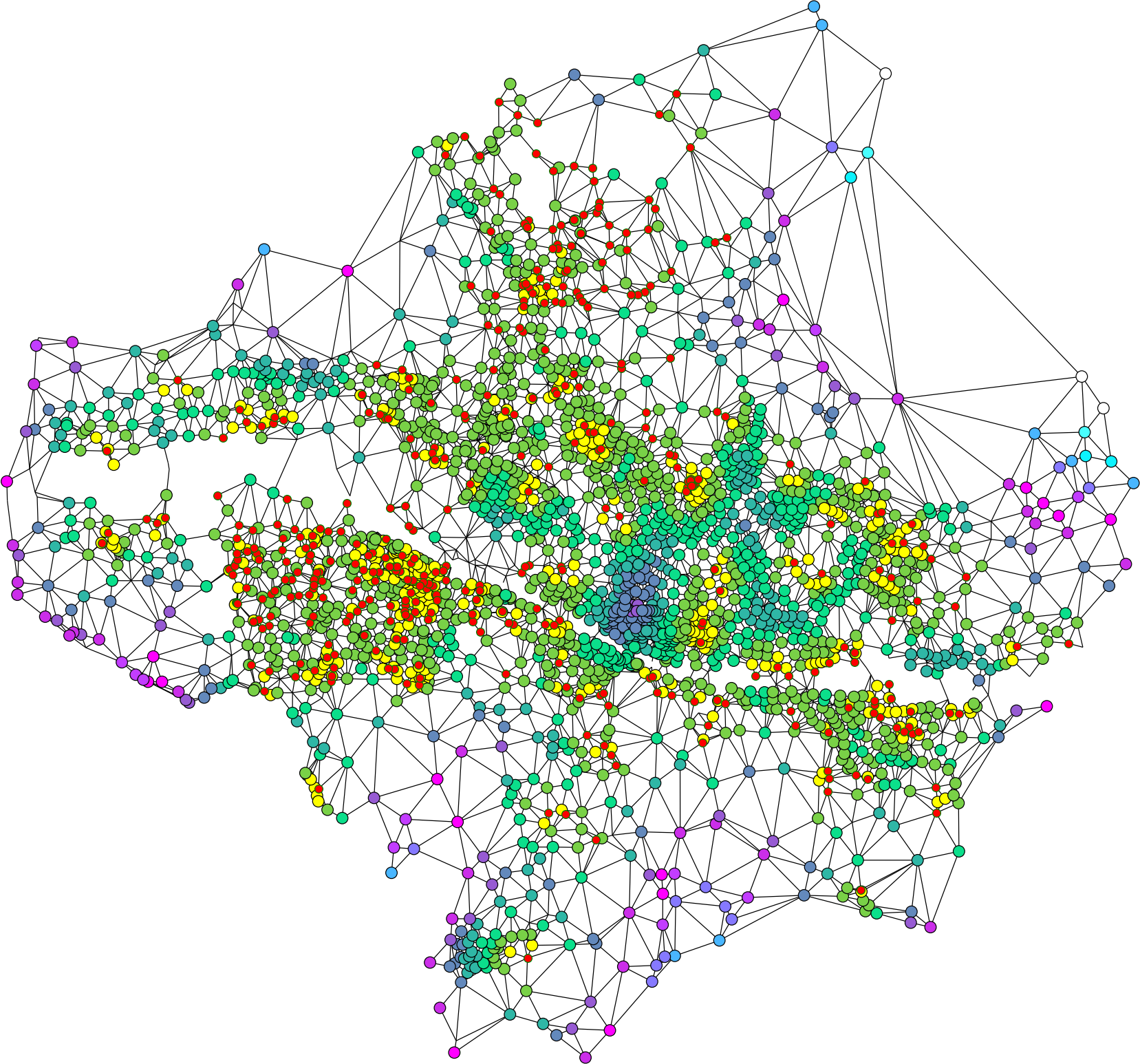}\label{fig:cent:avfarn:firenze}
        (b) Firenze\\
    \end{minipage}
    \hspace*{+1cm}
    \begin{minipage}[b]{0.33\textwidth}
        
         \label{fig:cent:mestre}
        \includegraphics[trim={0.5cm 0.1cm 0.5cm 0cm},clip,width=\textwidth]{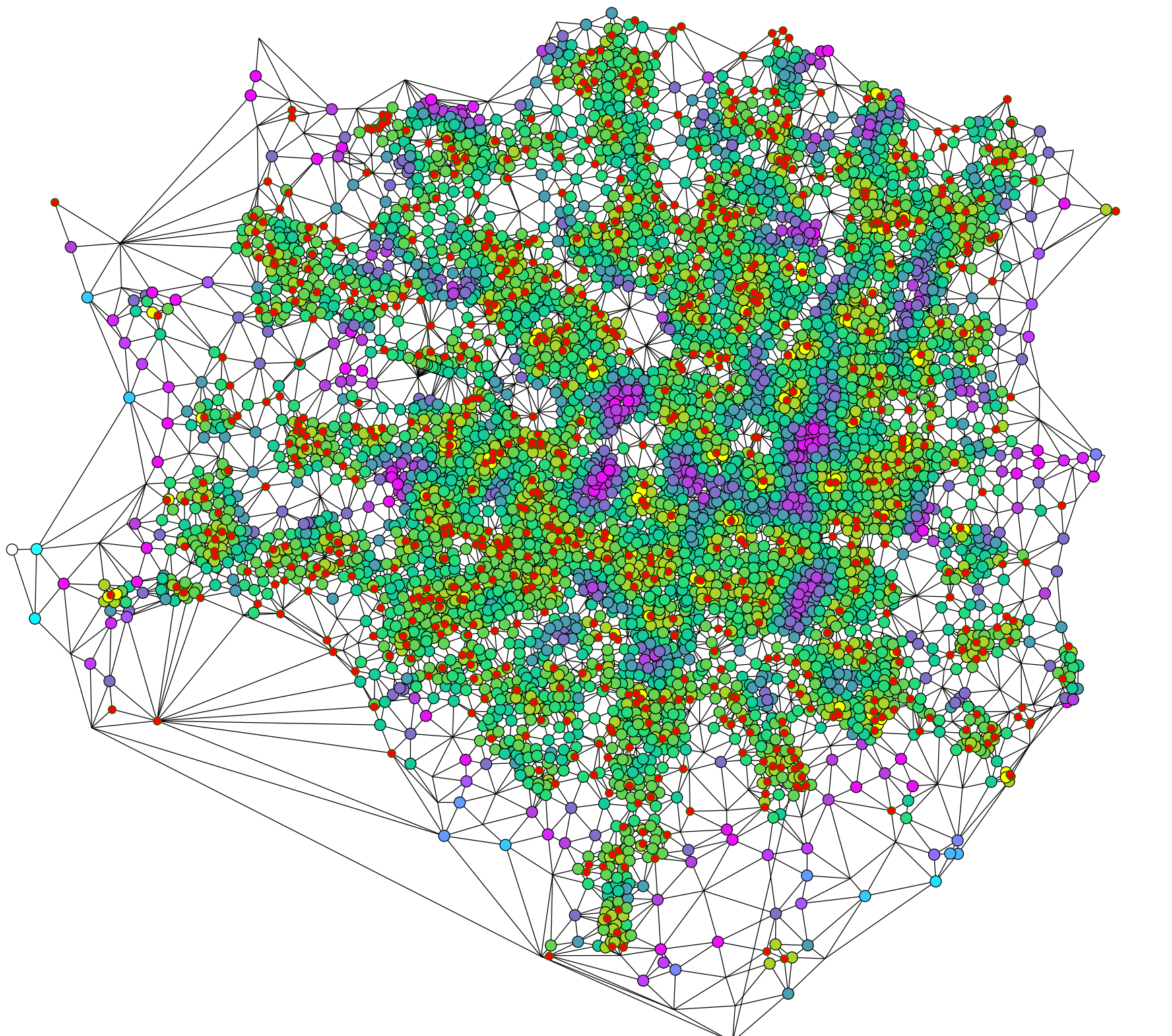}\label{fig:cent:avfarn:milan}
        (c) Milano\\
    \end{minipage}
    \raisebox{0.8cm}{ 
        \begin{minipage}[b]{0.04\textwidth}
            \includegraphics[width=1.5\textwidth,height=115pt]{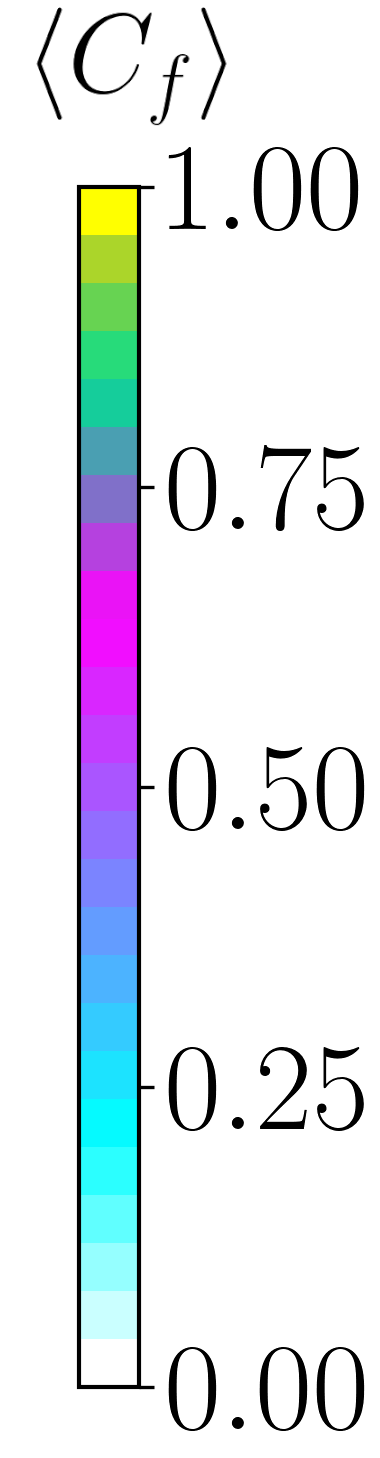}
        \end{minipage}
    }
    \caption{Average Farness centrality $\langle C_f \rangle$ mapped onto the respective spatial networks. PGA nodes are are shown (red). }\label{fig:avfarn}
\end{figure}
We show in Fig.~\ref{fig:avfarn} the depicted Average Farness following the formulation in Eq.~\eqref{eq:avgh} weight the accessibility by tempering the measure with the proximity of the block to nearby PGAs. Note that the different spatial scales of the three city clearly positively affect this measure and reveal the spatial organization of the PGAs across the city network. In detail, for Mestre, most of the blocks across the city exhibit an highly equitable accessibility to PGAs, when considering distance as the main characteristic of quality. We can explain  the spatial origin of this phenomenon from the organization of PGAs into small clusters close together, which allows the blocks to feel their proximity.
For Florence, on the other hand, this consideration fails when applied to its city center. Here, most of the high-centrality blocks lay in two large clusters on opposite banks of the river, corresponding to highly serviced spatial zones in terms of green areas. We put in relation this feature with the presence of environmental constraints on the planning of these areas, which appear naturally confined to only one part of the city.
For Milan, high quality blocks are concentrated near smaller PGA clusters. In this case, the spatial scale of the city, which is considerably larger than Florence and Mestre, affects the centrality. We measure that few blocks are highly served ($\langle C_f \rangle \sim 0.95$) while most retain a value around $\langle C_f \rangle \sim 0.75$. In essence, these observations reveal a kind of neighborhood planning at block-level PGAs in the city of Mestre and a rather different arrangement in the city of Florence, governed by morphological conditions. In the city of Milan, they reveal a quasi-uniform arrangement of PGAs although arge areas are missing green zones. This modified version of the Farness Centrality metric, better defined in Methods, improves the classical Farness Centrality used in this field. Indeed instead of computing centrality from a disk of infinite amplitude we now consider  the probability of observing low-cost paths heading to the respective PGAs. In fact, the latter is constrained by both the finite nature of the spatial network and the unphysical idea of assigning uniform weights to all paths to PGAs. However, this latter measure does not incorporate information about the area of the PGAs and considers these as equally contributing to the quality assessment, therefore avoiding the intrinsic scarcity that characterizes PGAs. Moreover, as anticipated, urban blocks are characterized by different population densities, and neither of these measures account for it.

\subsection{Average farness from occupancy distribution}
\begin{figure}[!ht]
    \centering
        \begin{minipage}[b]{0.45\textwidth}
                \centering
                \includegraphics[width=\textwidth]{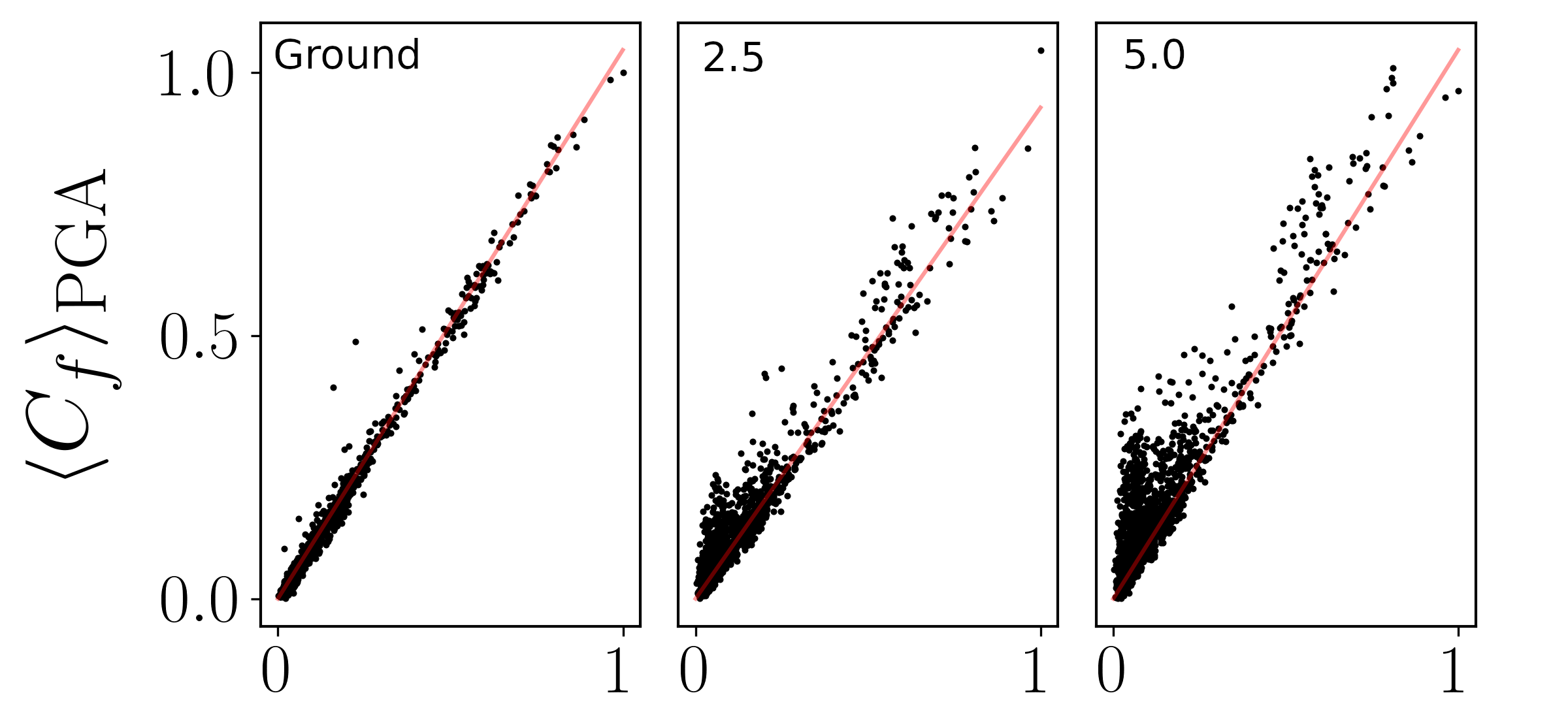}
                (a) Mestre\\
        \end{minipage}
        \begin{minipage}[b]{0.45\textwidth}
                \centering
                \includegraphics[width=\textwidth]{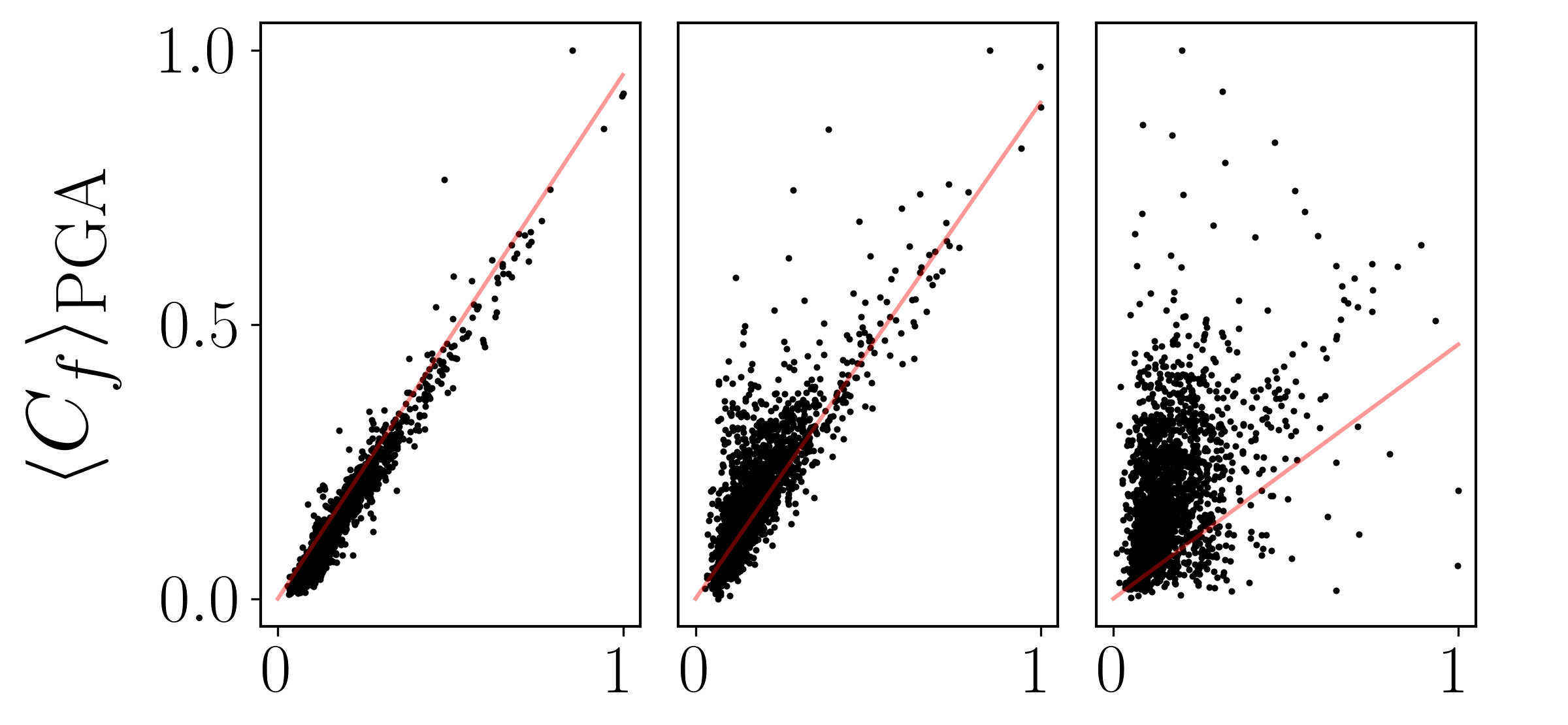}
                (b) Firenze\\
        \end{minipage}
        \begin{minipage}[b]{0.45\textwidth}
                \centering
                \includegraphics[width=\textwidth]{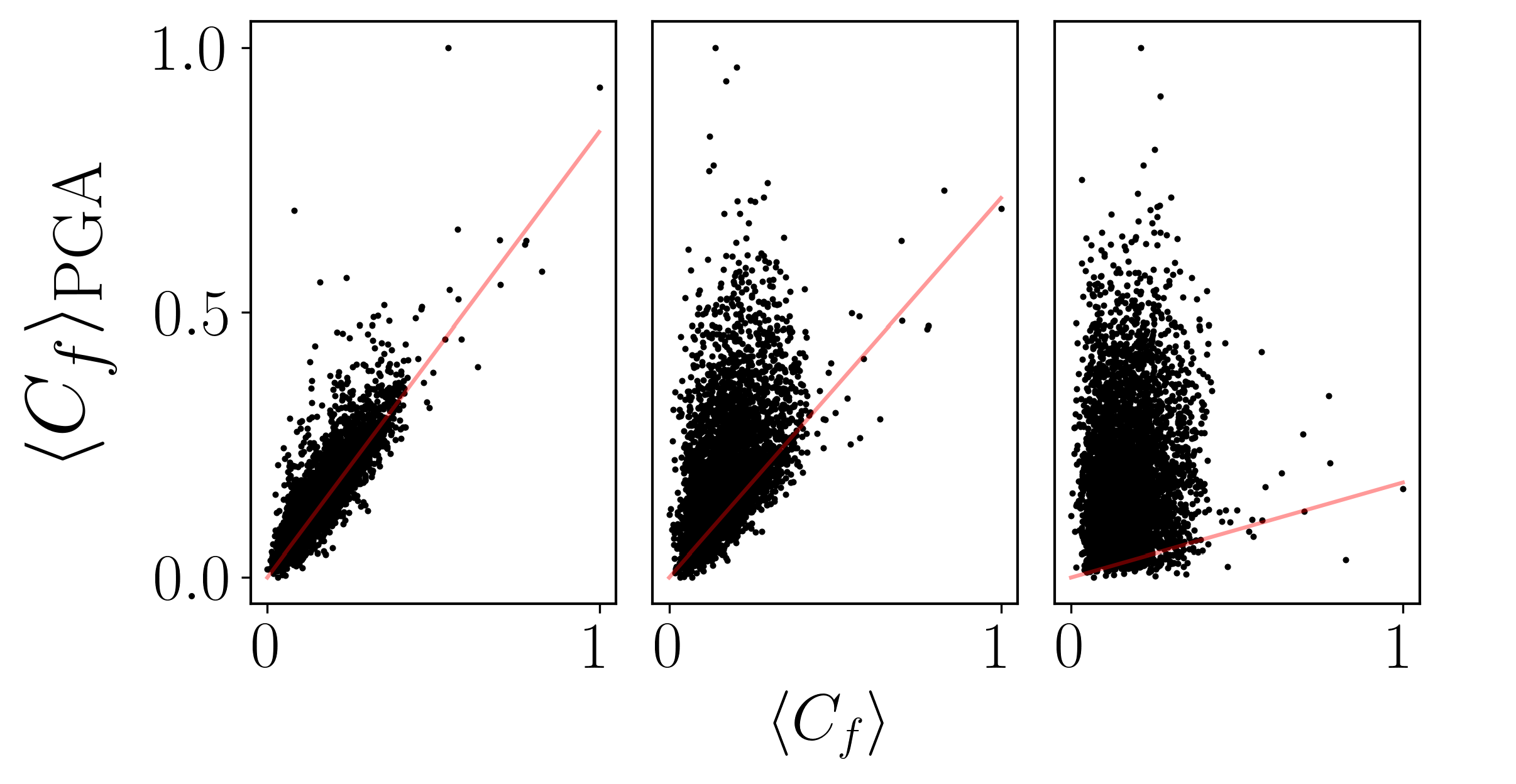}
                (c) Milano\\
        \end{minipage}
        \caption{Scatter plot of the agent-based Average Farness centrality vs the Average farness per block at different $f_{mq}$ (ground, 2.5, 5.0). Red line correspond to linear fit of the points.}\label{fig:corrcent}
\end{figure}

In this section, we report the results of the average farness from occupancy distribution obtained by the agent-based model. Before starting computer  simulations, we extract from the $\mathbf{Z^{\mathbf{PGA}}}$ (derived in Subsection \ref{subsec:avfarn}) the corresponding $\phi$ values for each block node within every city.
The agent-based simulations have been performed with an amount of $45000$ agents for each of the three cities, corresponding to $20\%$ of the total population in the case of Mestre, $4\%$ for Milan, and $12\%$ for Florence. Each city was simulated by setting $f_{mq}$ parameter to different values, including the 'ground' case in which the total capacity of PGA nodes is much larger than the number of agents. We set every agent to the same velocity of $1.33\frac{m}{s}$ and simulations were run for $5000$ steps, which is a sufficient amount of time to allow the set of agents to occupy their place. For agents, we set the maximum $E$ to $\SI{5000}{m}$ and the $\beta$ parameter was set to 1, in order to not further bias the sampling.
\begin{figure}[!ht]
    \centering
    \begin{minipage}[b]{0.35\textwidth}
        \includegraphics[trim={2cm 1.4cm 1.5cm 1.4cm},clip,width=\textwidth]{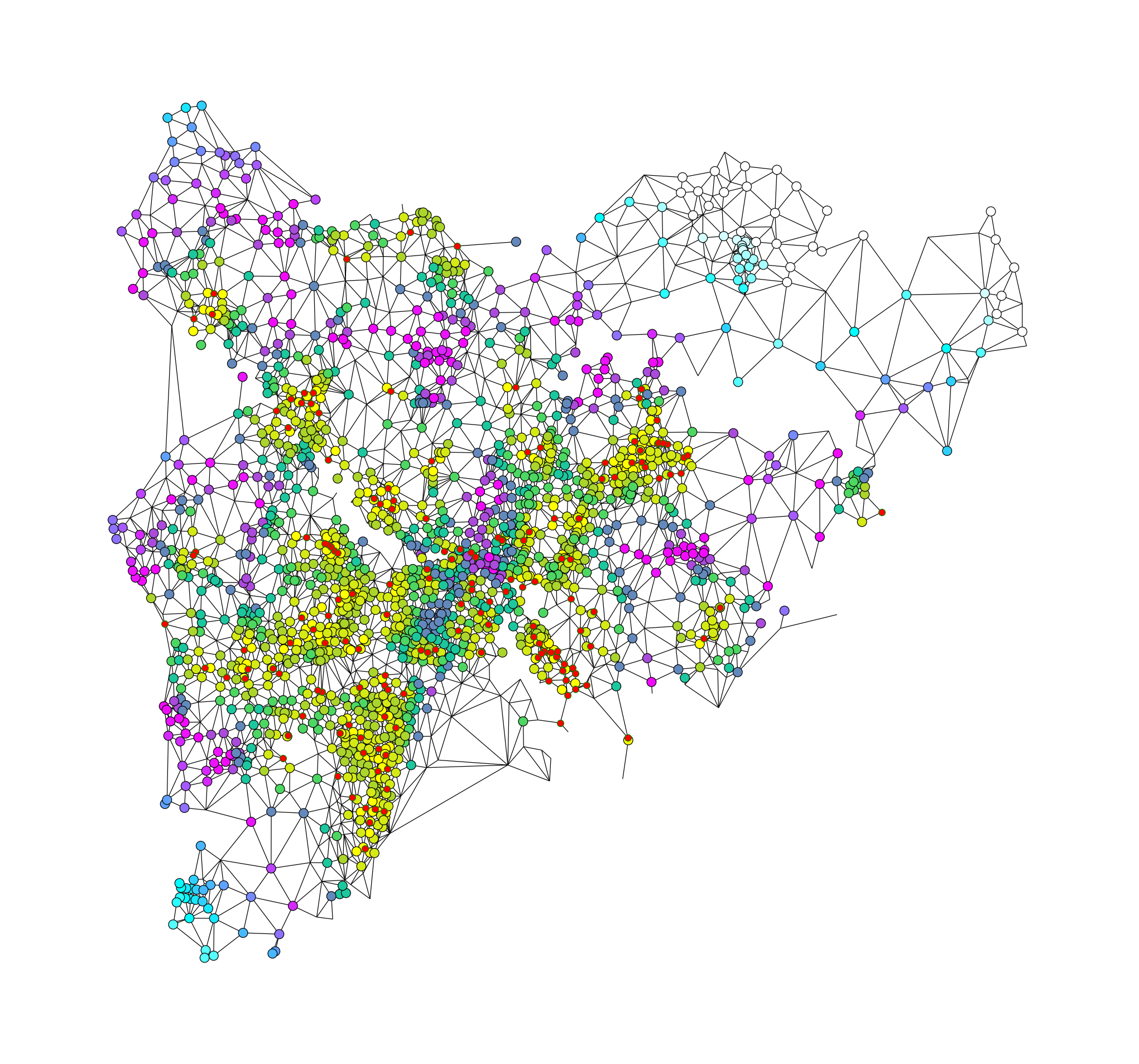}
        (a) Mestre\\
    \end{minipage}
    \begin{minipage}[b]{0.35\textwidth}
        \includegraphics[trim={1.5cm 1cm 1.3cm 1.8cm},clip,width=\textwidth]{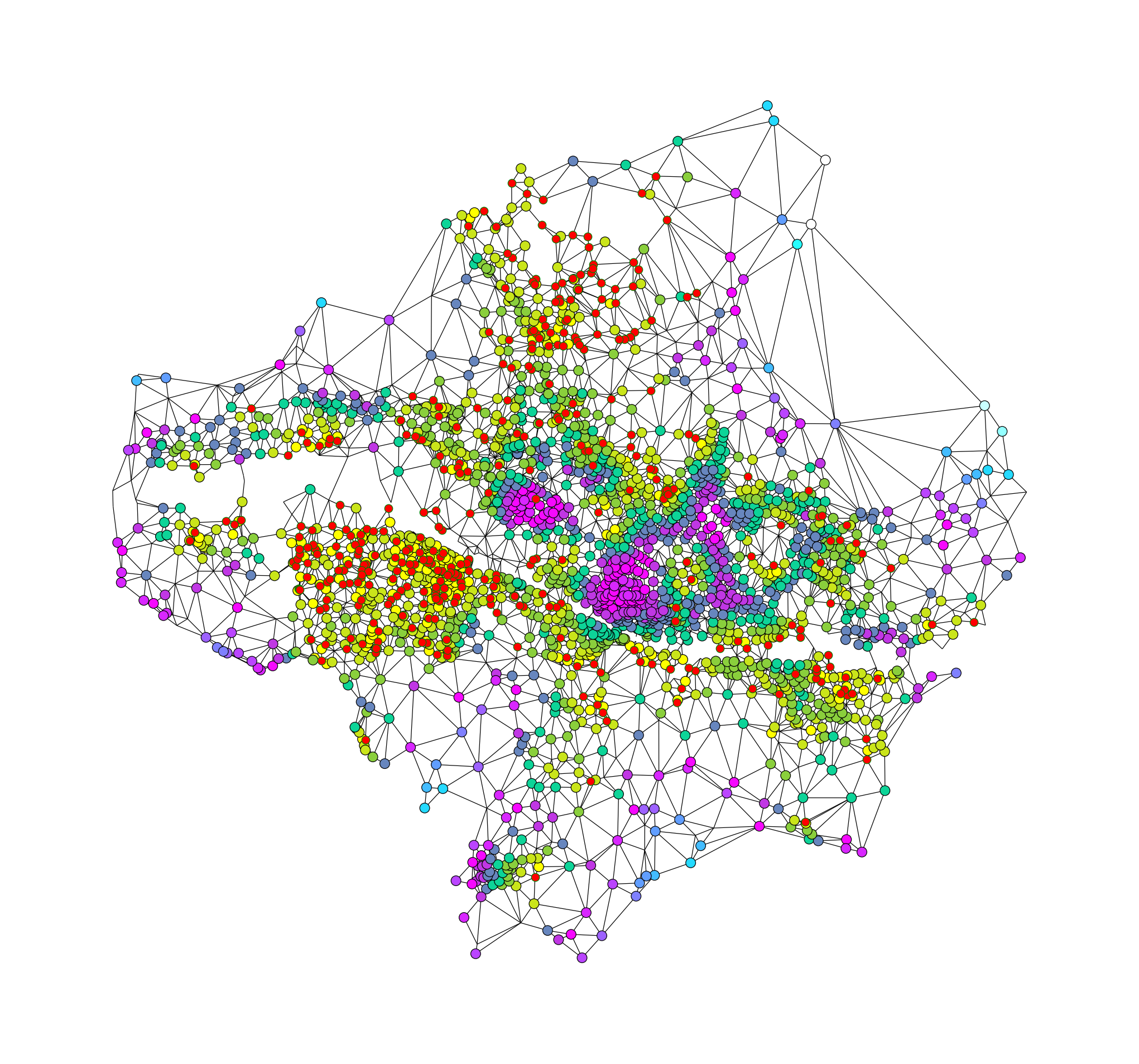}        
        (b) Firenze\\
    \end{minipage}
    \hspace*{+1cm}
    \begin{minipage}[b]{0.33\textwidth}
        \includegraphics[trim={2cm 2cm 1.8cm 1.4cm},clip,width=\textwidth]{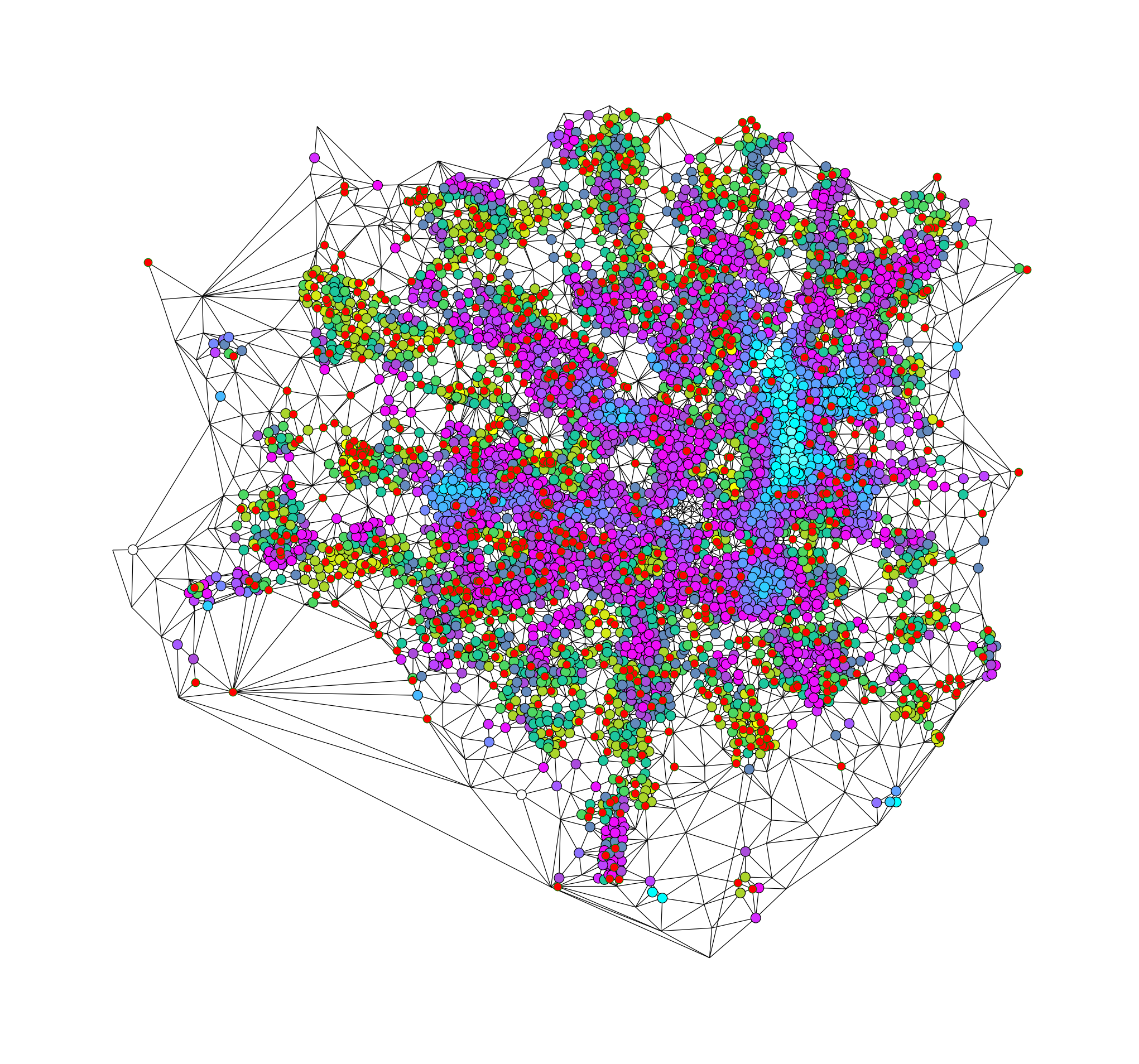}   
        (c) Milano\\
    \end{minipage}
    \raisebox{0.8cm}{ 
        \begin{minipage}[b]{0.04\textwidth}
            \includegraphics[width=1.5\textwidth,height=115pt]{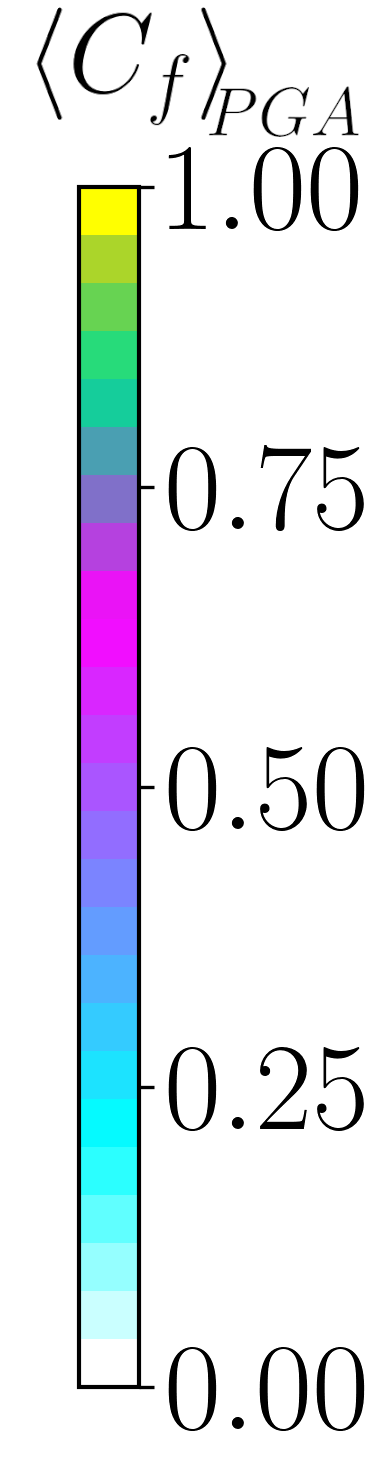}
        \end{minipage}
    }
    \caption{Average Farness centrality $\langle C_f \rangle_{PGA}$ from agent-based model mapped onto the respective spatial networks. PGA nodes are shown (red).}\label{fig:avfarnag}
\end{figure}
In essence, this metric resolves the lack of valuable information on both Farness Centrality and Average Farness centrality. 
In Fig.\ref{fig:avfarnag} the Average Farness centrality calculated as in Eq.\eqref{eq:avghag} is reported from the agent-based simulations described in Methods. Compared to Fig.\ref{fig:avfarn} it is possible to note a partial overlap and expansion of low-quality clusters for the three cities. For Mestre, as for Florence, it is possible to note the appearance of enlarged low-quality zones. While for Florence it was also evident from the Average Farness $\langle C_f \rangle$ the presence of local fragility given by the low presence of PGAs, for Mestre such fragility can be reported mainly for the modeled competition, where it appears clear that green areas are not sufficient to ensure equitable access for downtown residents. 
For Milan, the analysis reveals a more pronounced accessibility inequality.
From Fig.\ref{fig:avfarnag} it is possible to note a low-quality zone $\langle C_f \rangle_{\text{PGA}}\sim0.25$ (cyan) spanning the peripheral blocks of the city.
The peripheral belt of blocks within these centrality values matches low-quality clusters found in spatial studies \cite{transform2023access} which assume population density and PGAs total area as the driving factor of urban green quality.
The difference between the quality of blocks reported for the Average Farness and for the agent-based Average Farness can be thus understood from the non-cooperative competition of agents toward the PGAs induced both by the accessibility at different distances.
To explore to which extent agents compete for PGA access, the agent-based Average Farness centrality $\langle C_f\rangle_{\text{PGA}}$ is plotted versus the Average Farness $\langle C_f \rangle$ in Fig. \ref{fig:corrcent}.
\begin{figure}[!ht]
    \centering
        \begin{minipage}[b]{0.36\textwidth}
                \centering
                \includegraphics[width=\textwidth]{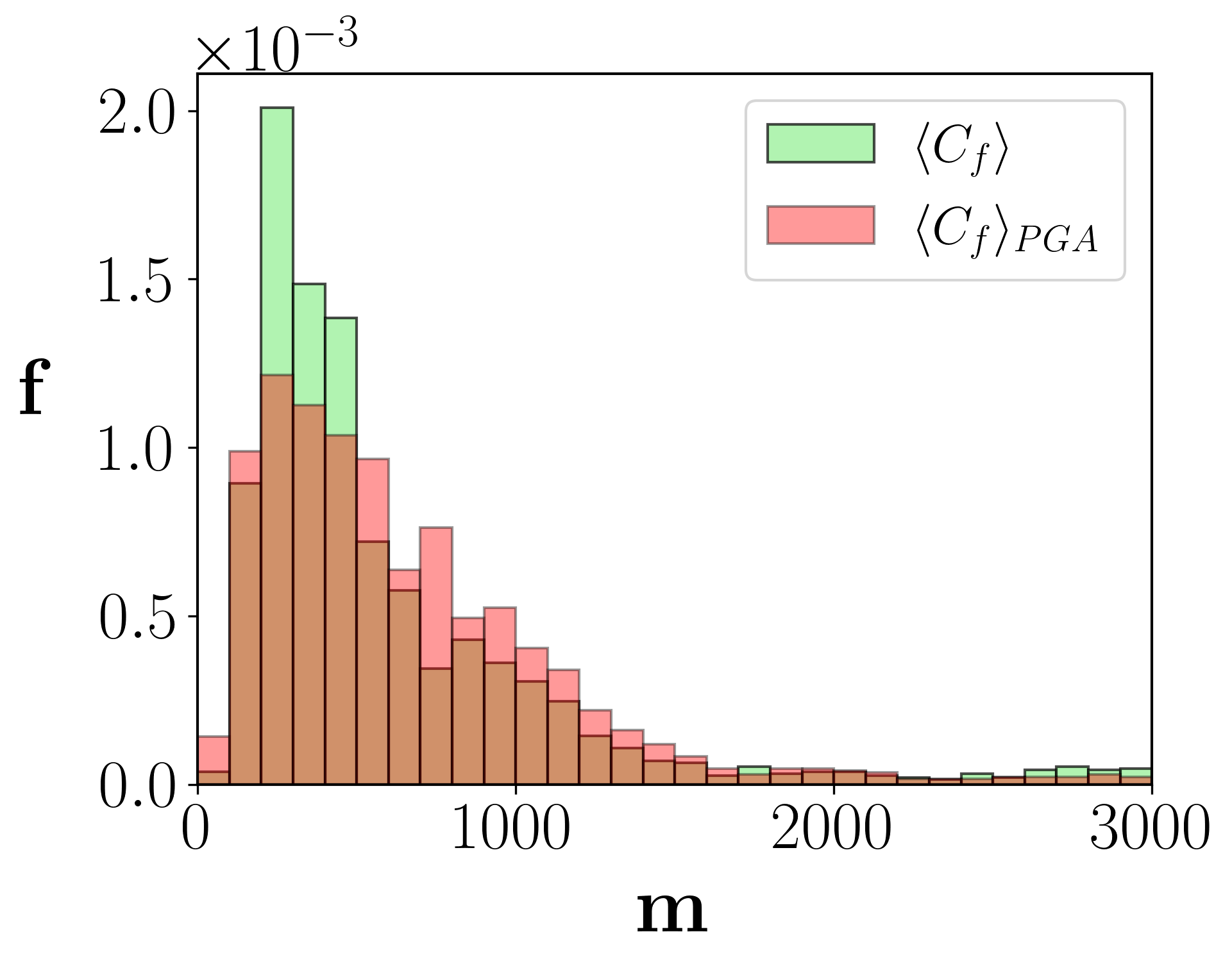} 
                (a) Mestre\\
        \end{minipage}
        \begin{minipage}[b]{0.36\textwidth}
                \centering
                \includegraphics[width=\textwidth]{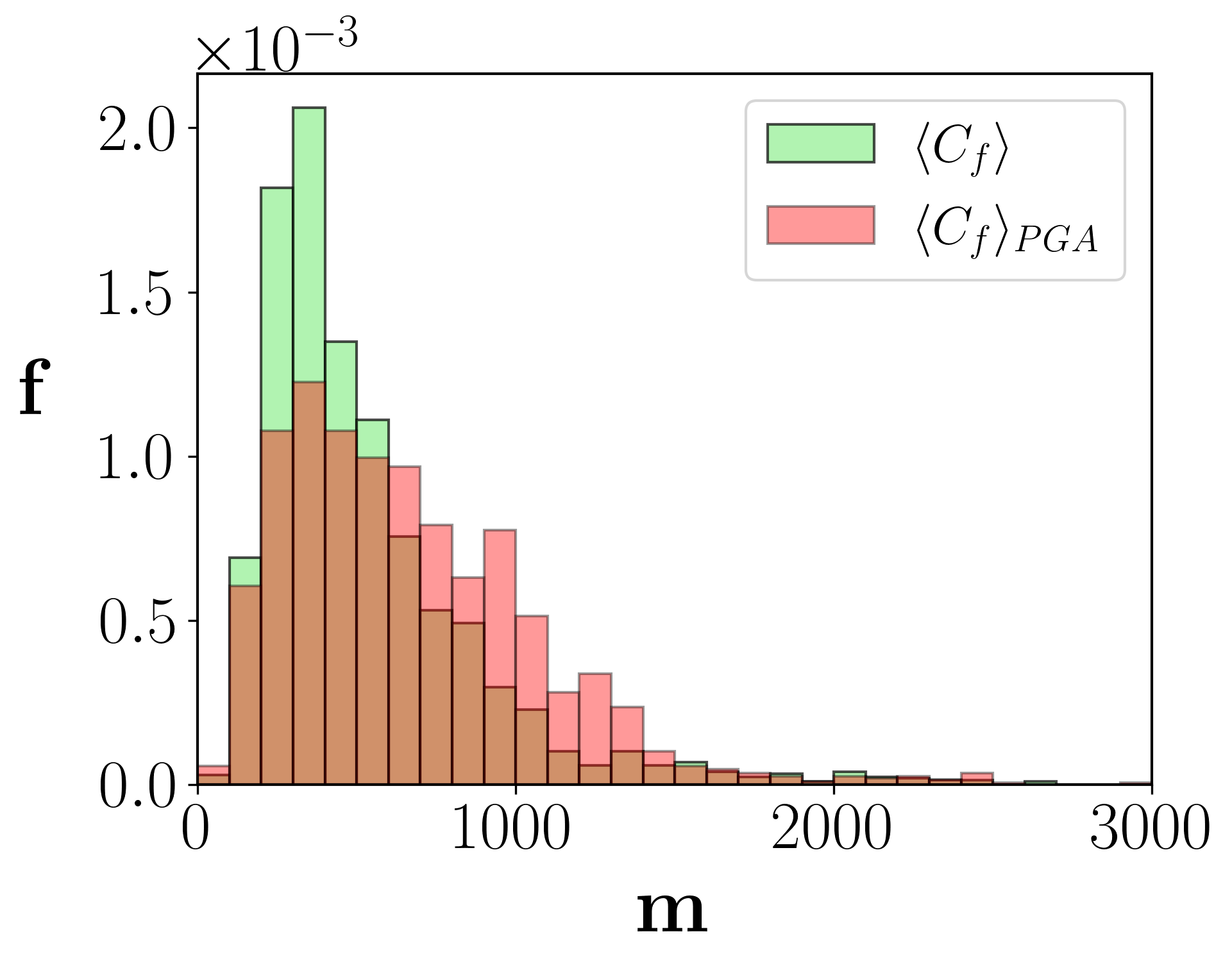} 
                (b) Firenze\\
        \end{minipage}
        \begin{minipage}[b]{0.35\textwidth}
                \centering
                \includegraphics[width=\textwidth]{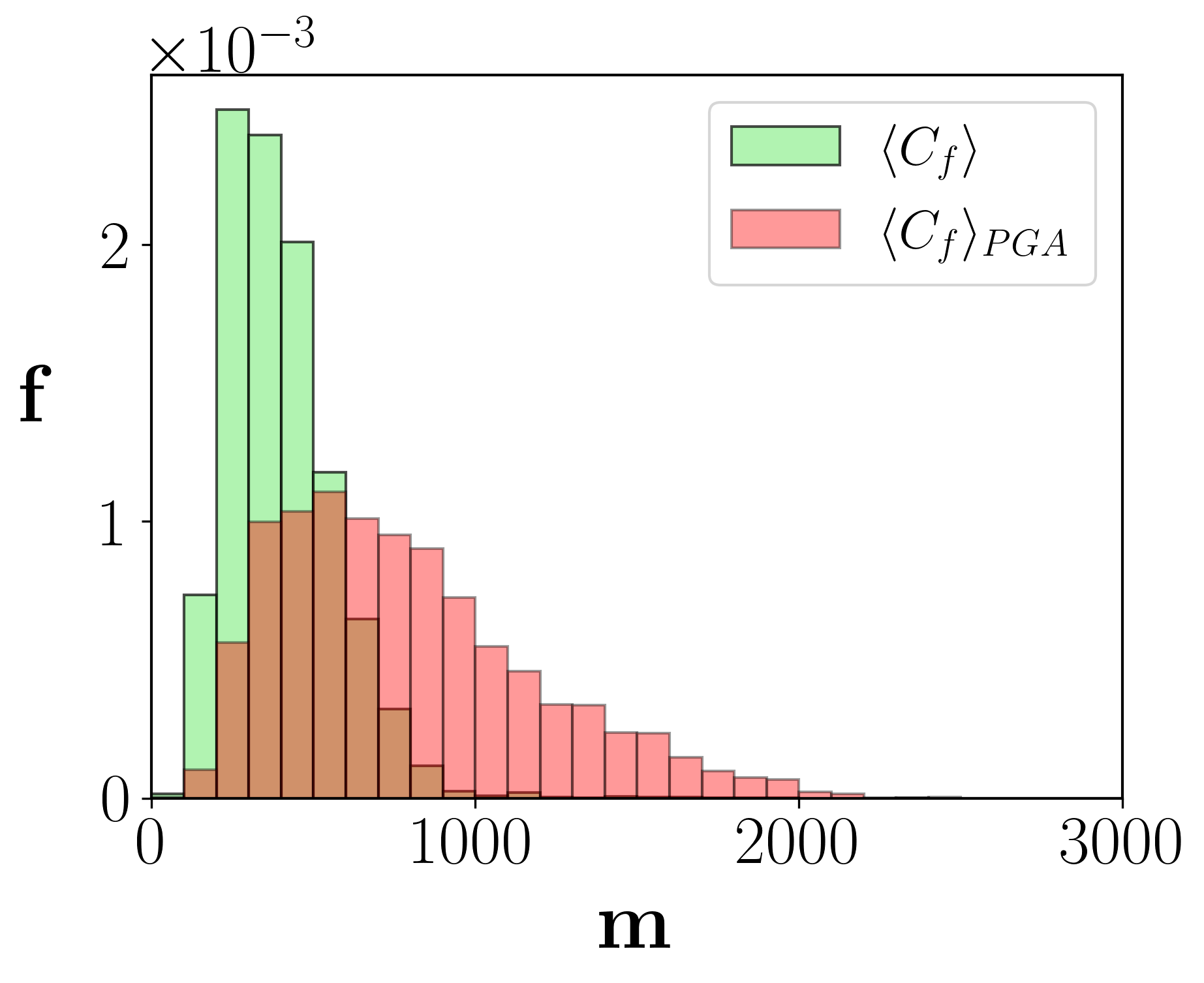} 
                (c) Milano\\
        \end{minipage}
        \caption{Histogram of unnormalized $\langle C_f \rangle_{\text{PGA}}$ and $\langle C_f \rangle$ per block expressed in meters at different distance bins $m$ ({\em{i.e.}} probability density of blocks that fall in the $i$-th bins). }\label{fig:histcent}
\end{figure} 
In particular, the figure shows a distinguishable pattern: at $f_{mq}=\text{ground}$, which corresponds to high PGA capacity, the Average Farness shows a linear correlation with agent-based one which is broken at increasing $f_{mq}$. First, this emphasizes that the assumption underlying the simulations method is verified because under high-capacity conditions agents sample paths proportional to the probability of accessing the nearest ones, thus recovering the Average Farness as a proportion of path probability. In the second instance, it reinforces the observation that, according to this model, the population gradient together with the increasing scarcity of green areas modifies the frequency with which agents are accepted by PGAs. 
Indeed, histograms reported in Fig.~\ref{fig:histcent} show how events considered 'rare,' ({\em{i.e.}} traveling long distances to access a PGA), become increasingly frequent as the area sought is constrained by the actual provision of green areas. This results in a distribution more shifted toward higher distance values; this effect is prominent in the case of Milan. 
This points out that PGAs become increasingly correlated with blocks at long distances. 
\subsection{Cumulative density as diffusion analysis}
\begin{figure}[!ht]
    \centering
     \includegraphics[width=0.44\textwidth]{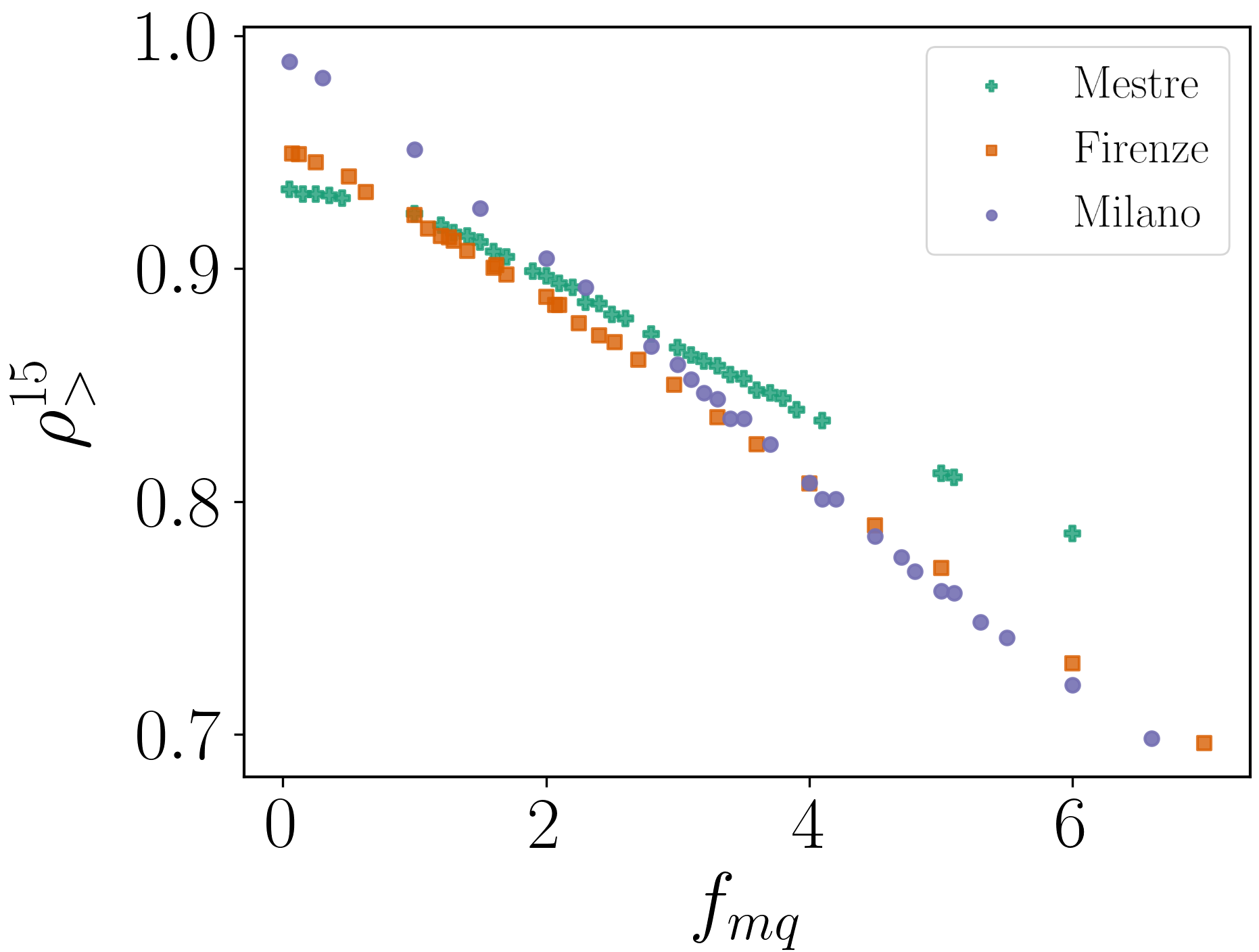}
     \caption{Fit of cumulative density  at the 15-minutes distance vs different $f_{mq}$. Fitted points correspond to the ones that intersect the 15-minutes distance (green line).}\label{fig:dens:fit}
\end{figure}

\begin{figure}[ht] 
\centering
    \begin{minipage}[b]{0.18\textwidth}
        
         \includegraphics[trim={0cm 0.3cm 0cm 0cm},clip,width=1\textwidth]{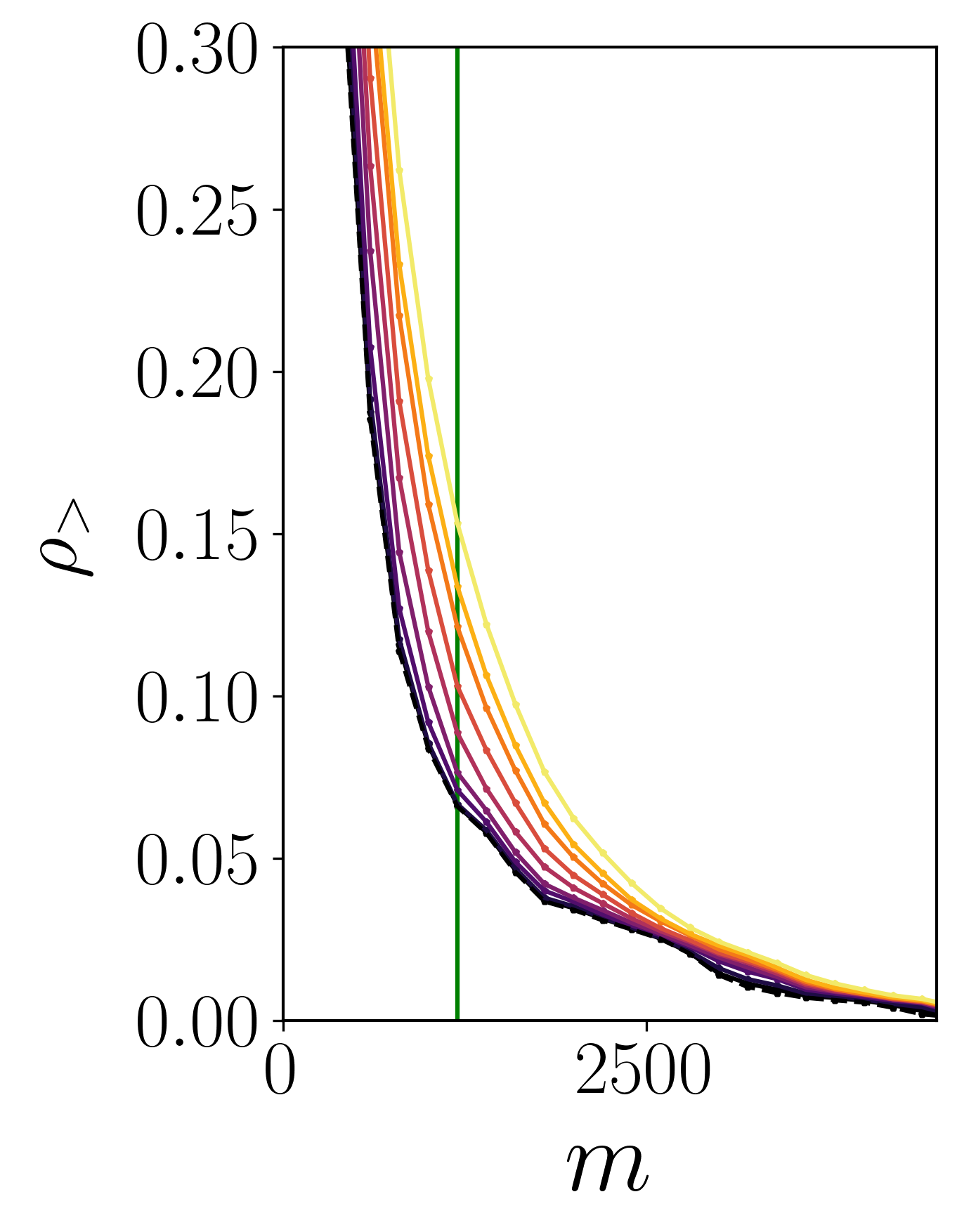}
         (a) Mestre\\
    \end{minipage}
    \begin{minipage}[b]{0.13\textwidth}
         \includegraphics[width=1\textwidth]{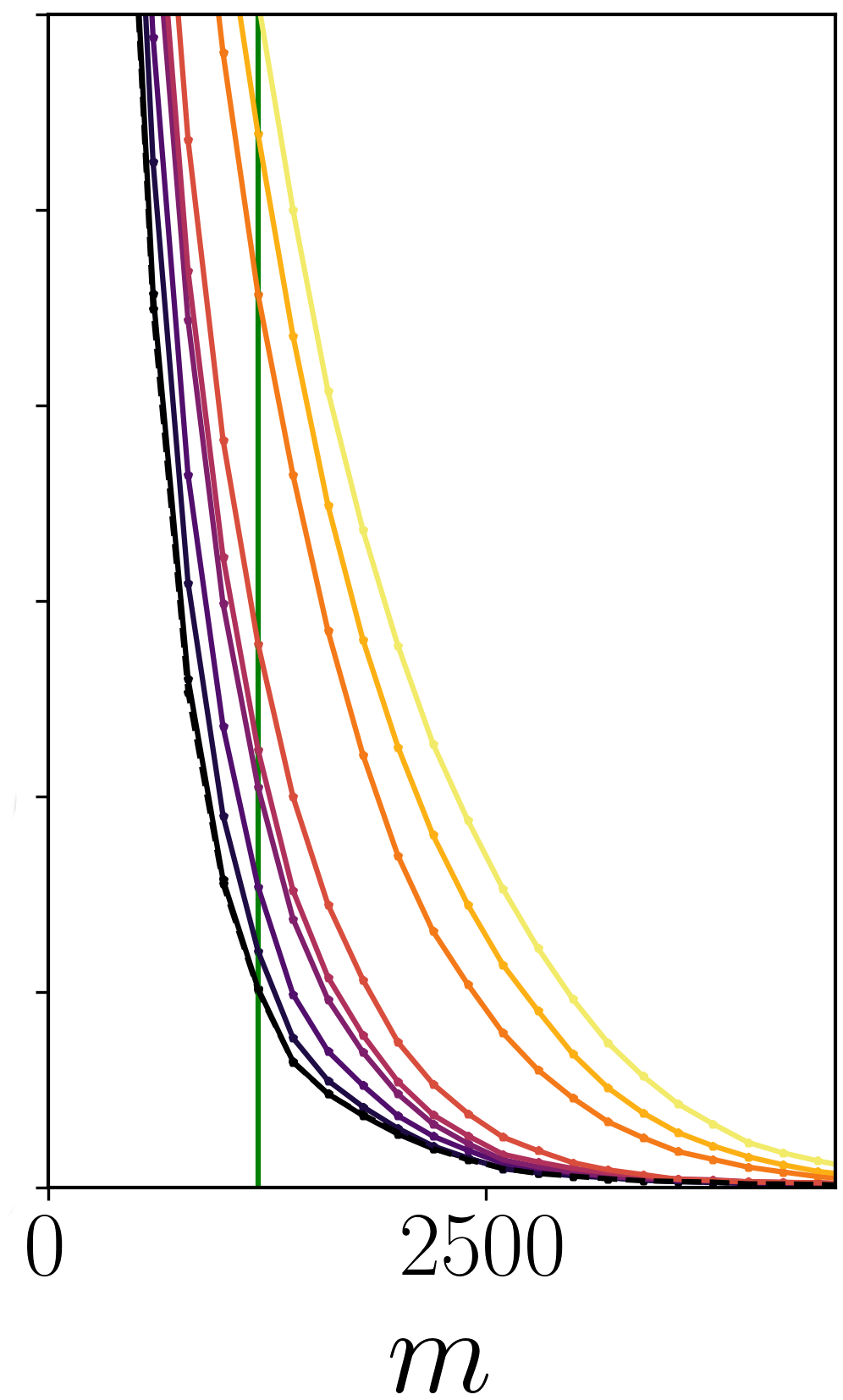}
        (b) Firenze\\
    \end{minipage}
    \begin{minipage}[b]{0.126\textwidth}
         \includegraphics[width=1\textwidth]{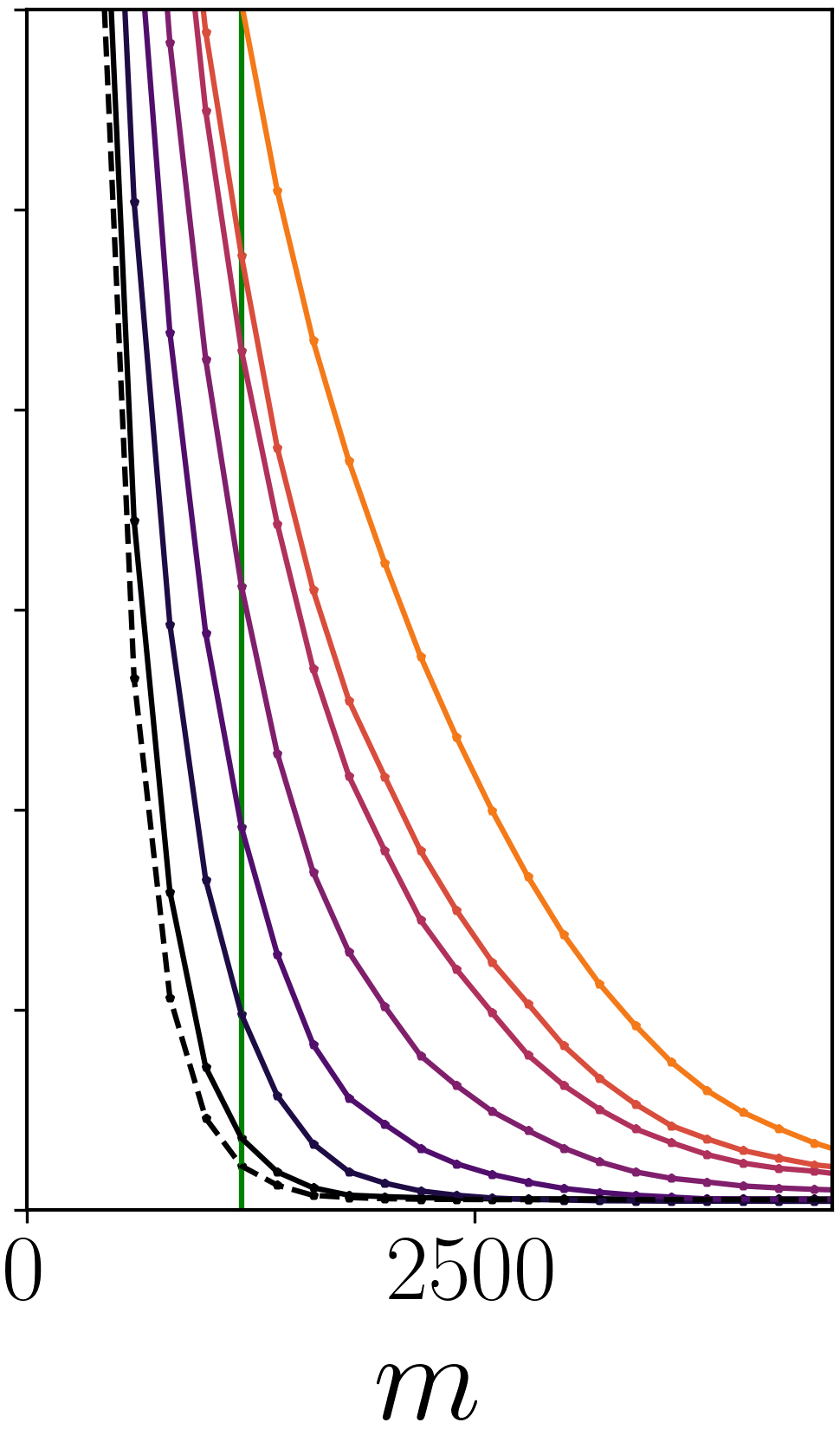}
         (c) Milano\\
    \end{minipage}
    \caption{ Cumulative density for the three cities at different $f_{mq}$.  }\label{fig:cumdensfit}
\end{figure}

We show in Fig.\ref{fig:cumdensfit} the Cumulative density calculated onto the previous simulated systems. t is possible to observe that, at increasing $f_{mq}$, the PGAs nodes source most of the agent's population at increasing distances. To understand how this property scales with respects to the 15-minutes distance, we show in Fig. \ref{fig:dens:fit} the fit of the data reported by the cumulative density at increasing $f_{mq}$. These plots suggest a linear relationship between the two quantities which decrease with a specific coefficient. The linearity follows the diffusion/adsorption mechanism, since the simulated copies of the system vary only by the parameter $f_{mq}$. The flux of agents on edges entering PGAs is proportional to the probability of observing a given path to them, and the fraction of agents accepted ({\em{i.e.}}, removed from the flux) depends on the actual occupancy of the designated PGA. A negative change in $f_{mq}$ thus produces a linear decrease in the fraction of accepted agents. The negative slope then indicates globally the degree of crowdedness present on PGAs walkable areas.
\hfill
Finally, we point out that as $f_{mq}$ increases, it is possible that a fraction of agents may not have reached an available PGA, thus slightly biasing the results. In this case, though, the simulations were controlled to minimize this problem. From our simulation we obtain a fraction less than $2\%$ of the total of agents still wandering at the end of the allowed exploration time.
\section{Conclusion}\label{sec:IV}
We presented a study on the usability of Public Green Areas (PGA) on spatial networks representation of cities. We obtain a quantitative measure of this feature by using centrality measures taken from network theory. First, we used the Farness Centrality from the classical point of view. A substantial modification was then introduced with the concept of Average Farness, by introducing optimal path distributions. To further characterize centrality through population difference, we also introduced a simple agent model, where agents compete in a non-cooperative manner for access to PGAs following low-cost paths toward these. Then we compared the centralities and we derived information regarding the effective distance of PGA to city zones.
As explained in the results, classical farness centrality is partially able to elucidate the quality of blocks with respect to PGA nodes, because these  paths are equally probable and therefore unrealistic  since distance is obviously positively correlated with an augmented energy spending in terms of walks. Consequently, we defined a more comprehensive measure of the likelihood of observing a given path to PGA's.
The probability of recovering a particular path given edge weights allows us to elucidate the spatial relation between blocks and PGA by escaping a strictly local description. In that sense we can then give a comparison between the Average Farness and the agent-based method.
We can compare statistically this metric with Average Farness $\langle C_f\rangle$ . In the latter, PGAs occupancy is dependent only on the first passage probability of random walks starting from a block, thus resulting in a uniform occupancy distribution and therefore inherently assumed to have infinite capacity. In the latter, with the introduction of a capacity constraint on PGA nodes and the optimization of paths by agents, the sampled occupancy distribution depends on the maximum occupancy of PGAs and on those paths corresponding to the second pass of agents in case they are rejected by a PGA. In other words, the Average Farness itself represents an intrinsic advantage/disadvantage given distances as spatial attributes, while the agent-based ones weights independently this intrinsic characteristic and the competition as a function of distance and green areas provision.
The latter metric represents an attempt to dynamically improve urban sustainability planning, where traditionally quality metrics are often prescribed based on static spatial attributes, which can fail to grasp the factors that affect PGA accessibility, actual use and perception. Nonetheless, the agent model can be further developed, in several directions, such as, for example: by considering, on one side, the socio-demographic attributes and needs that affect actual agents’capability and propension to access green spaces; but also, on the other, by including other PGA dynamic attributes (beyond the static spatial ones) that affect actual PGA agents use, such as PGA capacity to provide, for example, sociality, safety, quietness, shadow, perceived aesthetics or sense of freedom \cite{chiesi2022small}; by incorporating in the model that, as sociality is among the strongest drives to go to PGA form for most agents, PGA crowdedness can also be perceived as an incentive rather than an obstacle to attendance; and, last but not least, by addressing the fact that green spaces access does not only occur by walking, but also by other means of transport. The success of this analytical and modelling effort can produce useful data to inform PGA design, planning, policy making, and management, allowing a greater and balanced exposure of citizens to the beneficial effects of green spaces and promoting social justice and equity in the urban environment.
\section*{Acknowledgments}
We acknowledge enlightning discussion with Stefano Mancuso.
\appendix
\section{Farness centrality}
\begin{figure}[ht!]
    \centering
    \begin{minipage}[b]{0.3\textwidth}
        \includegraphics[trim={2cm 1cm 4.6cm 1cm},clip,width=1\textwidth]{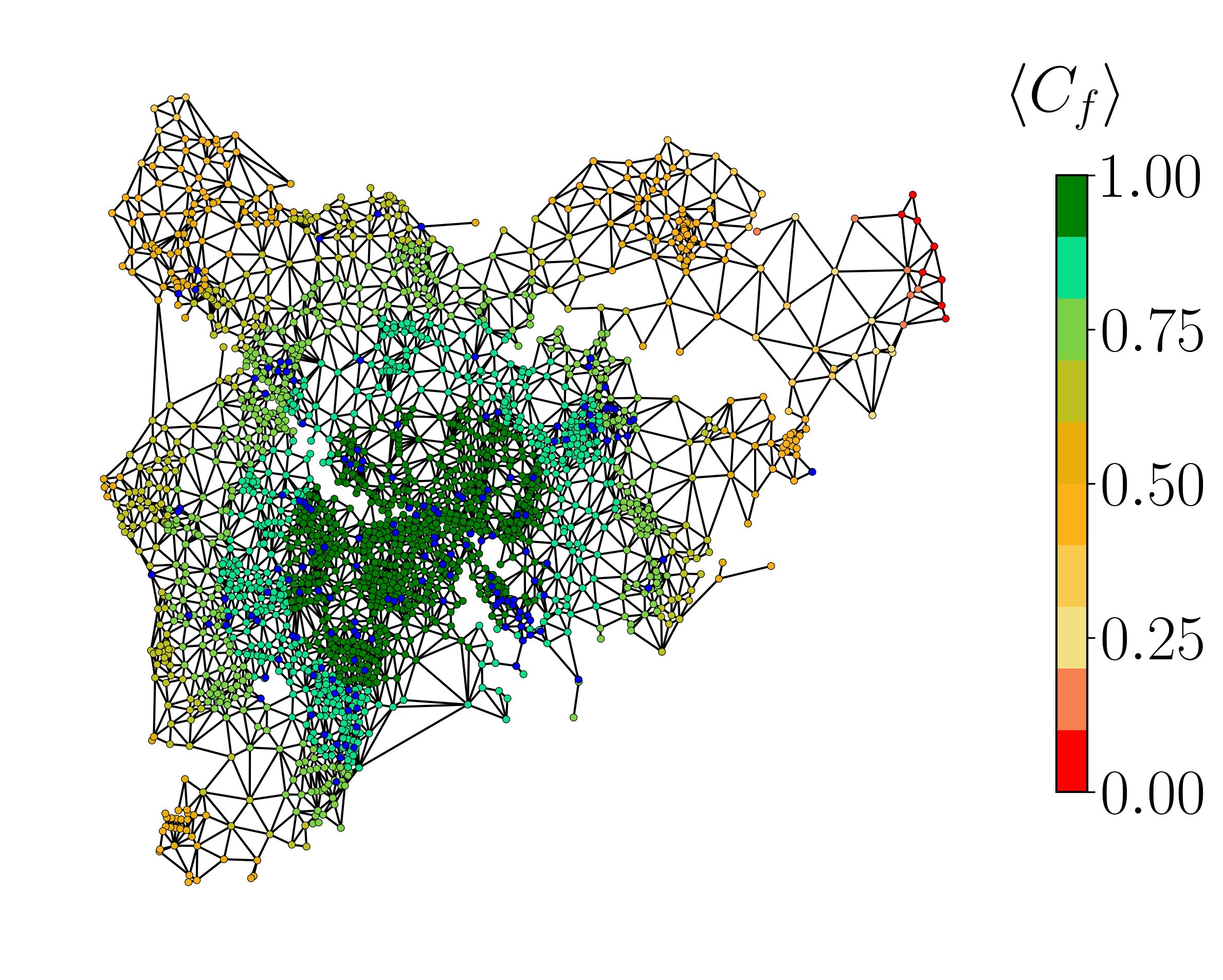}
        (a) Mestre\\
    \end{minipage}
    \begin{minipage}[b]{0.3\textwidth}
        \includegraphics[trim={1.4cm 1cm 1.7cm 1cm},clip,width=1\textwidth]{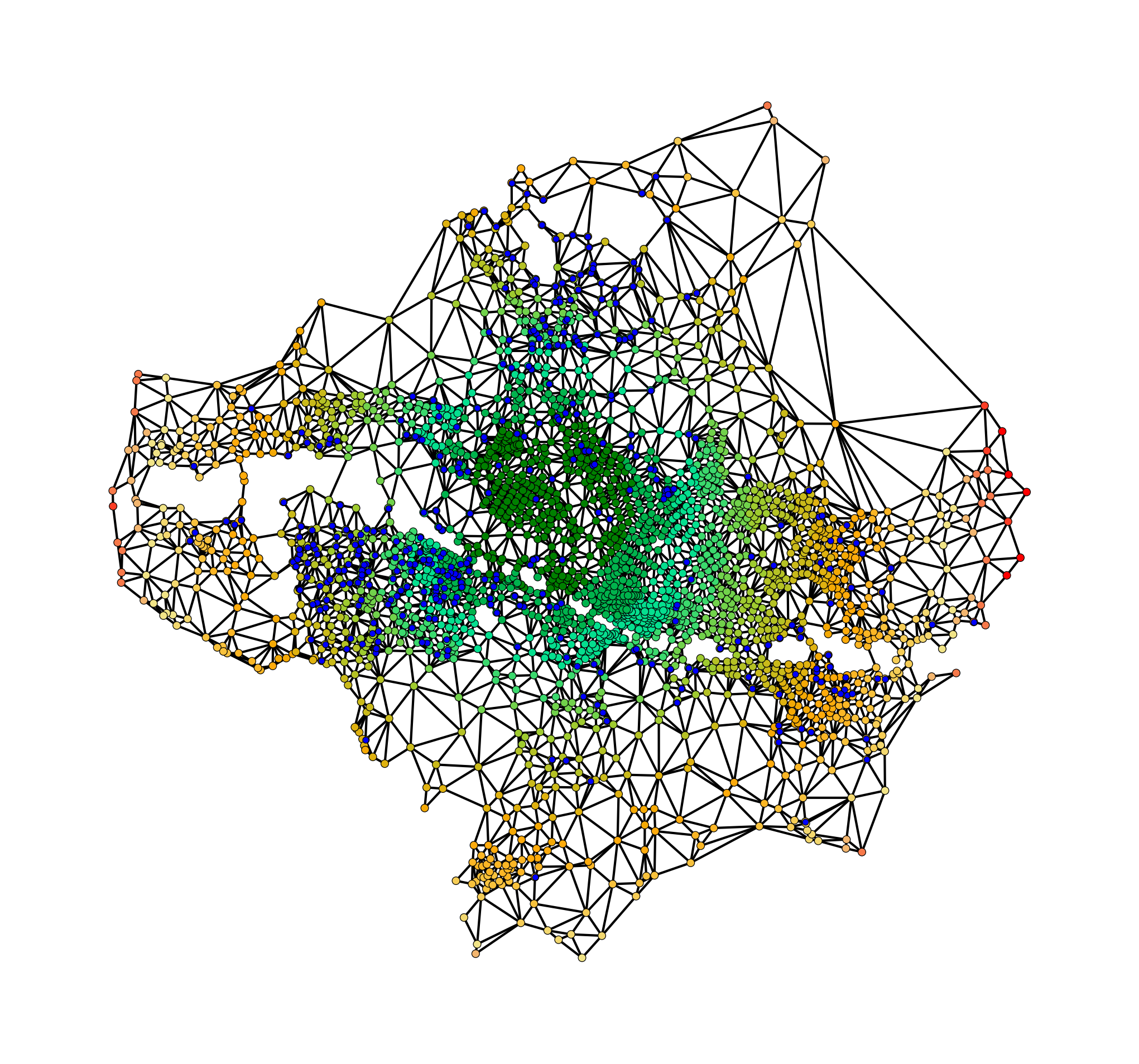}
        (b) Firenze\\
    \end{minipage}
    \hspace*{+1cm}
    \begin{minipage}[b]{0.3\textwidth}
        \includegraphics[width=\textwidth]{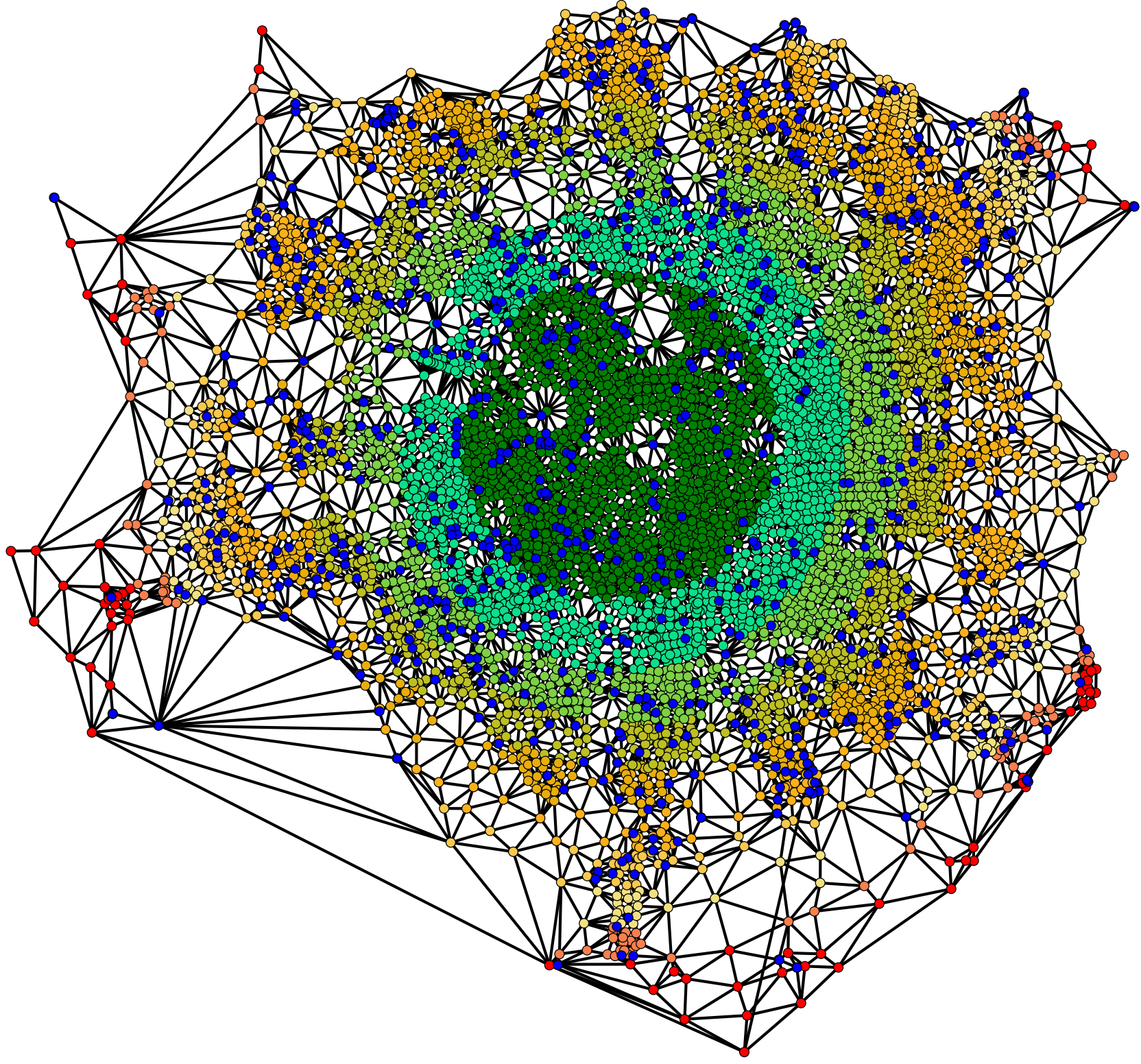}
        
        (c) Milano\\ 
    \end{minipage}
    \raisebox{0.8cm}{ 
        \begin{minipage}[b]{0.06\textwidth}
            \includegraphics[width=\textwidth,height=125pt]{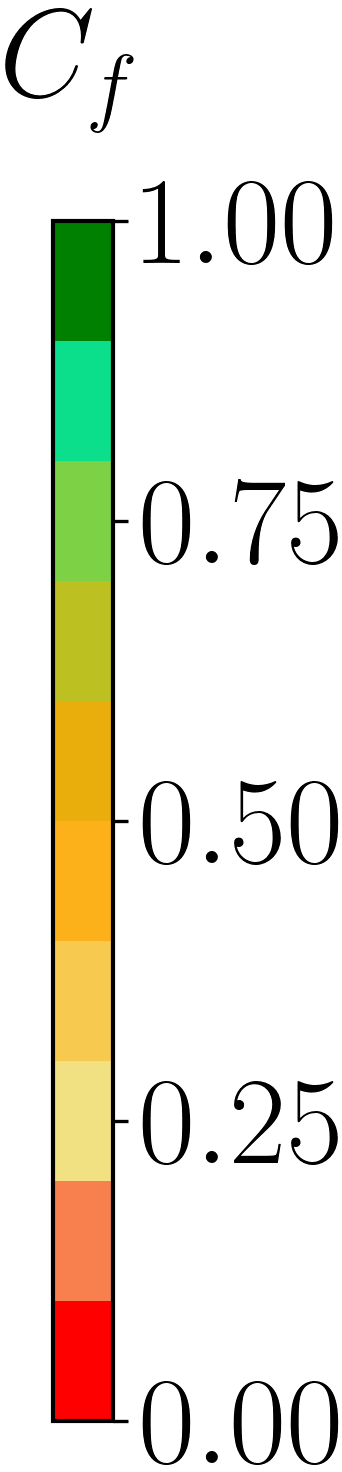}
           
        \end{minipage}
    }
    \caption{Farness centrality $C_f$ mapped onto the respective spatial networks with PGA nodes (blue).}\label{fig:farness}
\end{figure}
In Fig.\ref{fig:farness} the Farness centrality is reported. 
As can be seen from the Figure, Farness centrality uniquely identifies spatial zones with the minimum distances among the PGA node set, which is easy to. For Mestre and Milan, which have a sparse PGAs disposition, the area is identified near the city center while for Florence it identify zone near biggest PGA clusters.
Farness centrality tends to favor nodes that concentrate most network connections in different shells, often aligning with spatially central nodes. 
\section{Case studies in detail}
The study areas were obtained using the QGis tool \cite{QGISsoft}. Each area, representing a city and part of its conurbations, was obtained by intersecting the administrative boundaries of the cities with the census sections in the ISTAT dataset. A summary of the main numerical features of interest is given in Table \ref{tab:city_stats}.    
\begin{table}[ht!]
    \centering
    \caption{City Statistics}
    \label{tab:city_stats}
    \begin{tabular}{|l|c|c|c|}
        \hline
        & Extension($Km^2$) & Population($N$) & $N^\circ $ PGAs \\
        \hline
        Mestre  & 116 &  261.905 & 159  \\
        Firenze & 102 & 382.258 & 396  \\
        Milano  & 181 & 1.352.000& 753  \\
        \hline
    \end{tabular}
\end{table}
\\
\bibliography{parks}
\end{document}